\documentclass[aps,prb,twocolumn,secnumarabic,amsmath,amssymb,superscriptaddress,floatfix,eqsecnum]{revtex4-2}
\usepackage{amsthm}
\usepackage{amsmath}
\usepackage{amsfonts}
\usepackage{amssymb}
\usepackage{mathtools} 
\usepackage{bm} 
\usepackage{flafter}
\usepackage{slashed} 
\usepackage{color}
\usepackage{bm}
\usepackage{bbm}
\usepackage[breaklinks=true,colorlinks,citecolor=blue,linkcolor=blue,urlcolor=blue]{hyperref}
\usepackage{cases}
\usepackage{float}
\usepackage{url}
\usepackage{framed}
\usepackage{cancel,soul,ulem}
\usepackage{appendix}
\usepackage{graphicx}
\usepackage{multirow}
\usepackage{tikz}
\usepackage{braket}
\tikzset{>=latex} 
\usetikzlibrary{arrows.meta}
\usepackage{pgfplots} 
\pgfplotsset{compat = newest} 

\usepackage{ulem}

\begin{document}
\title{
Site-Order Optimization in the Density Matrix Renormalization Group \\ via Multi-Site Rearrangement
}

\author{Ryo Watanabe}
\affiliation{Graduate School of Engineering Science, The University of Osaka, 1-3 Machikaneyama, Toyonaka, Osaka 560-8531, Japan}
\author{Toshiya Hikihara}
\affiliation{Graduate School of Science and Technology, Gunma University, Kiryu, Gunma 376-8515, Japan}
\author{Hiroshi Ueda}
\affiliation{Center for Quantum Information and Quantum Biology, The University of Osaka, Toyonaka, 560-0043, Japan}
\affiliation{Computational Materials Science Research Team, RIKEN Center for Computational Science (R-CCS), Kobe, Hyogo 650-0047, Japan}

\begin{abstract}
In the approaches based on matrix-product states (MPSs), such as the density-matrix renormalization group (DMRG) method, the ordering of the sites crucially affects the computational accuracy.
We investigate the performance of an algorithm that searches for the optimal site order by iterative local site rearrangement.
We improve the algorithm by expanding the range of site rearrangement and apply it to a one-dimensional quantum Heisenberg model with random site permutation.
The results indicate that increasing the range of the site rearrangement significantly improves the computational accuracy of the DMRG method.
In particular, increasing the rearrangement range from two to three sites reduces the average relative error in the ground-state energy by 65\% to 94\% in the cases we tested.
We also discuss the computational cost of the algorithm and its application as a preprocessing for MPS-based calculations.
\end{abstract}

\date{\today}

\maketitle

\section{Introduction}\label{sec:Intro}

The density-matrix renormalization group (DMRG) method has been widely used and achieved great success in fields such as condensed-matter physics and quantum chemistry\cite{SCHOLLWOCK2011,Review2023}.
When first proposed by White\cite{White1992,White1993}, the DMRG was a variational method based on the matrix-product state (MPS) wave function\cite{OstlundR1995,RommerO1997}.
Thereafter, the high capability of MPS to represent the low-entangled states of quantum many-body systems has been recognized, and several powerful algorithms based on MPS have been developed, including the time-dependent DMRG\cite{WhiteF2004,DaleyKSV2004}, time-evolving block decimation\cite{Vidal2003,Vidal2004}, and MPS methods for finite temperatures\cite{White2009,StoudenmireW2010,IwakiSH2021}.
MPS has also been known as the tensor train and extensively applied to several topics in fields such as computational physics and machine learning\cite{TT_Review,MPS_ML,TT_DL,Shinaokaetal2023}.
Diverse and useful library tools have also been available to support MPS computations\cite{iTensor,iTensor03,10.1063/5.0180424,tenpy2024,van_damme_2025_17313329}.

MPS is a contraction of matrices (three-leg tensors, to be precise) arranged in a one-dimensional (1D) lattice.
In this sense, the structure of MPS as a tensor network is fixed.
However, MPS retains the freedom to select the order of physical sites in the 1D lattice.
This site order has been known to have a significant impact on the accuracy of MPS-based methods.
Many efforts have been devoted to identifying the optimal site order for a given Hamiltonian, and several schemes have been proposed\cite{ChanHG2002,LegezaS2003,MoritzHR2004,LegezaVPD2015}.

Recently, an intriguing algorithm for seeking the optimal site order has been proposed\cite{LiRYS2022}.
The basic idea of the algorithm is to sequentially improve the site order by incorporating a local two-site swap process into the DMRG calculation. 
The algorithm has been applied to several benchmark models and proven to be effective in improving computational accuracy.
However, for some models, the extent of accuracy improvement was modest, suggesting room for further refinement.
Incidentally, a similar algorithm of the structural optimization for tree-tensor networks has also been independently proposed\cite{Larsson2019,HikiharaUOHN2023,HikiharaUOHN2025,WatanabeMHU2026}.

A plausible explanation for why the algorithm of Ref.\ \cite{LiRYS2022} may not work well in some cases is that the site-order optimization is performed through the local exchanges between adjacent two sites.
Such an algorithm may get stuck in a local optimum in the site-order space from which the optimal site order can not be reached without a global rearrangement of sites.
A simple yet effective means to overcome the problem may be to implement a global site-order change by increasing the range of sites to be rearranged in the optimization process.

In this work, we have implemented the site rearrangement over a large region in the MPS containing $N_{\rm s}$ sites into the site-order optimization algorithm.
We have applied the algorithm to the spin-1/2 Heisenberg model in the 1D lattice with a random site permutation as a benchmark test.
We have thereby demonstrated that by increasing the rearrangement range $N_{\rm s}$, one can significantly improve the accuracy of the DMRG calculations.
Notably, increasing $N_{\rm s}$ from two to three reduces the random average of the relative error in the ground-state energy by 65\% to 94\% in the cases we tested.
The results provide a powerful preprosessing method for the site order to enhance the efficiency of MPS-based approaches.
The computational costs and techniques to reduce them have also been discussed.

The rest of the paper is organized as follows.
In Sec.\ \ref{sec:method}, we review the guideline to improve the site order, describe the algorithm of the site-order optimization, and discuss the computational cost.
In Sec.\ \ref{sec:NumRes}, we introduce the model to which the algorithm is applied and present the numerical results.
Section\ \ref{sec:conclusion} is devoted to a summary and concluding remarks.
Appendices\ \ref{app:store_reuse}-\ref{app:parallel} present the detailed derivations of the computational cost of the site-order optimization.

\section{Site-order optimization method}\label{sec:method}

\subsection{Effect of site order}\label{subsec:siteorder}

Here, we briefly review the importance of site order and the principle underlying the algorithm explored in this study.

\begin{figure}
\begin{center}
\includegraphics[width = 70 mm]{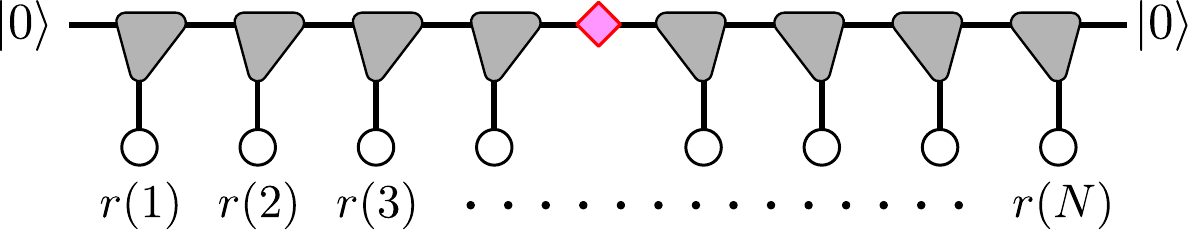}
\caption{
Matrix-Product State (MPS) for a system with $N$ sites.
Shaded triangles and red diamond represent the three-leg isometric tensors and the singular value diagonal matrix, respectively.
Open circles are the physical sites arranged in the order $\{ r(i)\}$, which is an arbitrary permutation of the site index $i= 1, 2, ..., N$.
The edge bonds are terminated by a trivial boundary vector $|0 \rangle$, which represents the dimension-one auxiliary index at the ends of the MPS.
}
\label{fig:MPS}
\end{center}
\end{figure}

MPS is constructed by taking a contraction of three-leg tensors arranged in one dimension (Fig.\ \ref{fig:MPS}).
Each three-leg tensor is assigned to a physical site.
The arrangement of the physical sites is generally arbitrary, and we suppose that the physical sites are arranged in the order of $\{ r(i); i=1,2,...,N\}$, which is a permutation of $\{1, 2, ..., N\}$.
Such a MPS is written as
\begin{eqnarray}
&&\sum_{\sigma_1,...,\sigma_N} \sum_{\alpha_0,...,\alpha_{N+1}}
A[r(1)]^{\sigma_1}_{\alpha_0 \alpha_1}
\cdots A[r(l)]^{\sigma_l}_{\alpha_{l-1} \alpha_l} 
\nonumber \\
&&~~~\times D_{\alpha_l \alpha_{l+1}}
B[r(l+1)]^{\sigma_{l+1}}_{\alpha_{l+1} \alpha_{l+2}} 
\cdots B[r(N)]^{\sigma_N}_{\alpha_N \alpha_{N+1}}
\nonumber \\
&&~~~~\times \delta_{\alpha_0,0} \delta_{\alpha_{N+1},0}|\sigma_1, \sigma_2, ..., \sigma_N \rangle,
\label{eq:MPS}
\end{eqnarray}
where $A[r(i)]$ and $B[r(i)]$ are three-leg tensors assigned to the physical site $r(i)$, $D$ is a singluar value diagonal matrix, and $\delta_{i,j}$ is the Kronecker delta.
Here, we have taken the mixed canonical form\cite{SCHOLLWOCK2011}, where the three-leg tensors $A$ and $B$, which are called isometries, satisfy the orthonormal condition,
\begin{eqnarray}
&&\sum_{\sigma_i,\alpha_i} \bar{A}[r(i)]^{\sigma_i}_{\alpha_i \alpha_{i+1}}
A[r(i)]^{\sigma_i}_{\alpha_i \alpha'_{i+1}}
=\delta_{\alpha_{i+1},\alpha'_{i+1}},
\label{eq:orthogonality_A} \\
&&\sum_{\sigma_i,\alpha_{i+1}} \bar{B}[r(i)]^{\sigma_i}_{\alpha_i \alpha_{i+1}}
B[r(i)]^{\sigma_i}_{\alpha'_i \alpha_{i+1}}
=\delta_{\alpha_i,\alpha'_i},
\label{eq:orthogonality_B}
\end{eqnarray}
where $\bar{A}$ and $\bar{B}$ are complex conjugates of $A$ and $B$, respectively.
The position of the singular value matrix, $l \in [1,N]$, is arbitrary.
The indices $\alpha_i$ connect neighboring isometries (or an isometry and the singular value matrix) and are referred to as auxiliary bonds. Their dimension is constrained by an upper bound $\chi$, known as the bond dimension.

In approaches based on MPS, we employ MPS in order to accurately represent the wave function of a quantum state by maximizing the fidelity.
A characteristic of MPS relevant for that purpose is that cutting any auxiliary bond splits the system into the left and right subsystems.
Therefore, to accurately represent a quantum state by MPS, each auxiliary bond must precisely express the entanglement between the subsystems connected by the bond.
However, due to the upper bound on the dimension of auxiliary bonds in MPS, the entanglement that the bonds can carry is also upper-bounded by a finite value.
Then, if the entanglement that an auxiliary bond must represent exceeds the upper bound, the excess amount of the entanglement causes a loss of accuracy.
To minimize this loss, the entanglement loaded on each auxiliary bond should be kept as small as possible.

\begin{figure}
\begin{center}
\includegraphics[width = 60 mm]{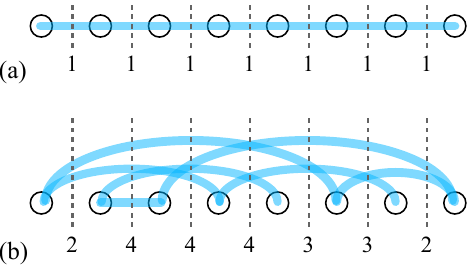}
\caption{
Schematic pictures of states with a single-stroke entanglement structure.
The blue line connects the sites that are strongly entangled.
Vertical dotted lines represent a cut of the system, and the numbers indicated for each dotted line denote the number of blue lines across the cut.
}
\label{fig:EEstructure}
\end{center}
\end{figure}

The importance of the site order is clear from the argument above.
Consider, as an example, a quantum state with an entanglement structure where the physical sites are connected in a single stroke, such as those shown in Fig.\ \ref{fig:EEstructure}.
In such a state, when the system is split into left and right subsystems, the entanglement between them is, roughly speaking, proportional to the number of blue lines across the two parts.
Therefore, it is obvious that the site order shown in Fig.\ \ref{fig:EEstructure}(a), where the sites are arranged according to the 1D entanglement structure, is optimal to minimize the entanglement on the auxiliary bonds.
In contrast, arranging the sites irregularly, say, as shown in Fig.\ \ref{fig:EEstructure}(b), increases the entanglement on the auxiliary bonds, causing a decrease in the precision of the MPS representation of the state with a finite bond dimension.

\begin{figure}
\begin{center}
\includegraphics[width = 60 mm]{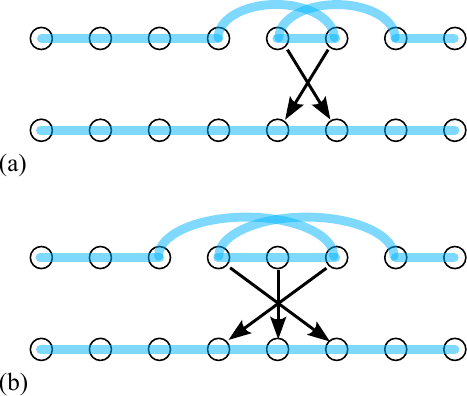}
\caption{
Schematic pictures of site order optimization for a state with a single-stroke entanglement structure.
(a) State for which a two-site swap can achieve the optimal site order.
(b) State for which a three-site rearrangement is required to reach the optimal site order.
}
\label{fig:rearrange}
\end{center}
\end{figure}

The site order optimization algorithm proposed in Ref.\ \cite{LiRYS2022} searches for the optimal site order by iteratively performing the rearrangement of two adjacent sites.
Here, the range of the site rearrangement crucially affects the efficiency of the optimization.
Let us consider a state with the entanglement structure shown in the upper panel of Fig.\ \ref{fig:rearrange}(a).
In this case, only swapping the two sites indicated in the figure achieves the optimal site order.
In contrast, if the site order is as shown in the upper panel of Fig.\ \ref{fig:rearrange}(b), swapping two adjacent sites does not reduce the total entanglement on the auxiliary bonds (the sum of the number of blue lines across the cuts)\cite{note_EEreduction}.
Therefore, in such a situation of Fig.\ \ref{fig:rearrange}(b), the optimization involving only the two-site swap may get stuck.
This problem can be solved by expanding the range of the site rearrangement.
Specifically, for the case in Fig.\ \ref{fig:rearrange}(b), expanding the range to include three adjacent sites enables a smooth approach to the optimal site order.
We examine the impact of this expansion of the range of site rearrangement on the computational accuracy in Sec.\ \ref{sec:NumRes}.

\subsection{Algorithm}\label{subsec:algorithm}

The site-order optimization algorithm that we study consists of two parts.
The first part is to update the isometries in the MPS, and the second part is to update the site order.
By iterating these calculations alternately, the algorithm optimizes the site order.

In the first part, we perform the variational calculation for a given model to update the isometries in the MPS with a fixed site order.
This calculation is equivalent to a standard DMRG.

In the second part, we optimize the site order so as to minimize the entanglement for the MPS wave function obtained in the first part.
The rearrangement of the sites is conducted through the fusion and decomposition processes\cite{ShiDV2006} as follows.

\begin{figure}
\begin{center}
\includegraphics[width = 60 mm]{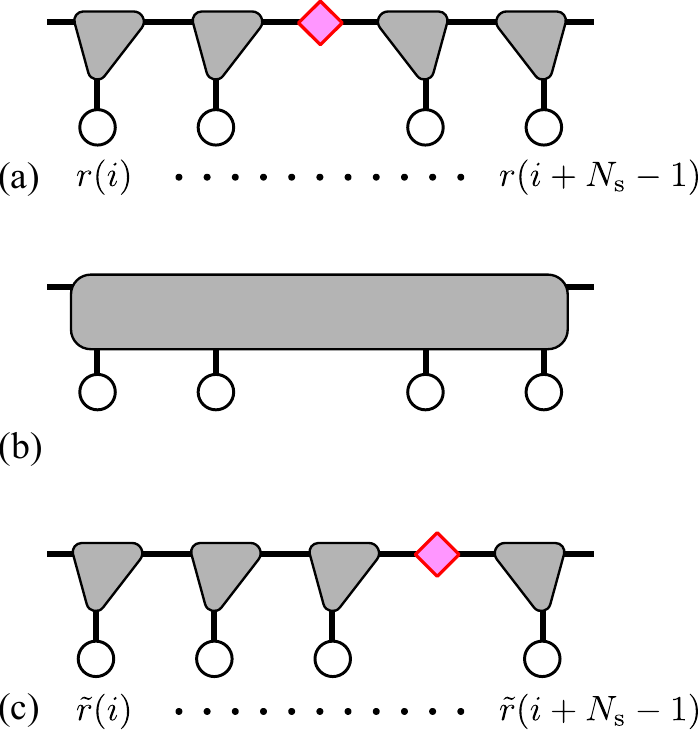}
\caption{
(a) Part of MPS focused at a step in the site-order optimization, including $N_{\rm s}$ isometries and a singular value matrix.
The indices of the physical sites in the region are $\{ r(i), ..., r(i+N_{\rm s}-1) \}$.
(b) Contraction of the isometries and singular value matrix in the focused region.
(c) Updated part of MPS in the focused region where the site order is modified to the optimized one, $\{ \tilde{r}(i), ..., \tilde{r}(i+N_{\rm s}-1)\}$, which is a permutation of $\{ r(i), ..., r(i+N_{\rm s}-1) \}$.
}
\label{fig:algorithm}
\end{center}
\end{figure}

Let us focus on a region of the MPS consisting of $N_{\rm s}$ adjacent isometries and a singular value matrix [Fig.\ \ref{fig:algorithm}(a)].
We take their contraction to generate an $(N_{\rm s}+2)$-leg tensor [Fig.\ \ref{fig:algorithm}(b)].
Then, we repeatedly perform singular-value decompositions (SVDs) to decompose the $(N_{\rm s}+2)$-leg tensor into a product of $N_{\rm s}$ isometries and a singular value matrix [Fig.\ \ref{fig:algorithm}(c)].
The SVD sequence follows the standard MPS/DMRG tensor factorization procedure: 
for a detailed exposition, see, {\it e.g.}, Fig.\ 5 of Ref.~\cite{SCHOLLWOCK2011}.
The crucial point here is that the singular values $\{ D_\alpha \}$ obtained at every decomposition step depend on the site order, i.e., the order in which physical sites are extracted one by one during the sequential SVDs;
the entanglement entropy on the auxiliary bond, $- \sum_\alpha (D_\alpha)^2 \log (D_\alpha)^2$, also depends on the site order accordingly.
In the algorithm, we compute the sum of the entanglement entropies for each of the $N_{\rm s}!$ candidates of site order and adopt the one that yields the smallest sum as the optimal site order.
(See Sec.\ \ref{subsec:comp_cost} and Appendices\ \ref{app:store_reuse}-\ref{app:parallel} for the details of the computation.) 
After that, we update the MPS by replacing the tensors in the focused region with those obtained by the SVDs for the optimal site order, and proceed to the next step to treat the region shifted by one isometry.
We perform this local site-order rearrangement iteratively by sweeping the entire MPS.

\begin{table*}
\caption{
Algorithm of the site-order optimization.
}
\label{tab:algorithm}
\begin{center}
\begin{tabular}{ll}
\hline
\hline
1. &Perform the standard DMRG with a fixed site order to update the isometries and singular value matrix.\\
2. &Perform the rearrangement of sites \\
   & (2-i) Take a contraction of adjacent $N_{\rm s}$ isometries and a singular value matrix in the focused region [Fig.\ \ref{fig:algorithm}(a)] \\
   &~~~~~~~~to make a $(N_{\rm s}+2)$-leg tensor [Fig.\ \ref{fig:algorithm}(b)].\\
   & (2-ii) 
Conduct SVDs to obtain the entanglement entropies on the auxiliary bonds for the possible $N_{\rm s}!$ site orders.\\
   & (2-iii) Identify the local optimal site order that minimizes the sum of the entanglement entropies on auxiliary bonds.\\
   & (2-iv) Update the MPS in the focused region according to the optimized site order.\\
   & (2-v) Iterate processes from (2-i) to (2-iv) by sweeping the entire MPS.\\
3. &Iterate processes 1 and 2 alternately until the site order and quantities to be studied converge. \\
\hline
\hline
\end{tabular}
\end{center}
\end{table*}

In our calculation, we have performed the DMRG calculation of the first part until the variational energy converges or a specific number of sweeps is reached.
We have then performed the second part for a single sweep (a round trip) of the entire MPS.
These two calculations have been iterated alternately until both the energy and site order converge\cite{convergence}.
Table\ \ref{tab:algorithm} summarizes the algorithm.

\subsection{Computational cost}\label{subsec:comp_cost}

\begin{table}
\caption{
Upper bound of the computational cost for the SVD process to evaluate the entanglement entropies for $N_{\rm s}!$ candidates of possible site order.
(a) Method to store and reuse intermediate tensors and (b) Method to use the table of entanglement.
Scaling on the range of site rearrangement $N_{\rm s}$, bond dimension $\chi$, and the degrees of freedom of physical site $d$, is presented.
Here, ``Truncation'' refers to the procedure in which a full SVD is performed at every decomposition step and then the bond dimension is truncated to $\chi$.
In contrast, ``Truncation*'' refers to the use of a partial SVD that computes only $\chi$-largest singular values and vectors without generating the full spectrum.
}
\label{tab:comp_cost}
\begin{center}
\begin{tabular}{cc}
\hline
\hline
\multicolumn{2}{l}{(a) Method to store and reuse intermediate tensors} \\
\hline
No truncation & $\mathcal{O}(N_{\rm s}!\chi^3  e^{d}d^{N_{\rm s}})$  \\
Truncation    & $\mathcal{O}(N_{\rm s}!\chi^3  e^{d}d^{2})$  \\
\hline
\hline
\multicolumn{2}{l}{(b) Method to use table of entanglement} \\
\hline
No truncation & $\mathcal{O}(2^{N_s}\chi^{3}d^{N_s+\lfloor N_s/2\rfloor})$ \\
Truncation*    & $\mathcal{O}(2^{N_s}\chi^{3}d^{N_s})$ \\
\hline
\hline
\end{tabular}
\end{center}
\end{table}

The process with the highest computational cost in the site-order optimization algorithm is the SVDs to evaluate the entanglement entropies for $N_{\rm s}!$ candidates of the site order in the focused region of MPS.
The straightforward implementation of this process is to perform SVDs to separate isometries one by one from the one end of the $(N_{\rm s}+2)$-leg tensor for each of the $N_{\rm s}!$ site orders independently.
The resulting computational cost is $\mathcal{O}(N_s!\chi^{3} d^{N_s+\lfloor N_s/2 \rfloor})
$, where $\chi$ is the bond dimension of the MPS and $d$ is the number of degrees of freedom of each physical site.
Here, we assume that no truncation is performed on the singular values during the sequential SVDs.
This computational cost is significantly larger than that of the standard DMRG with a fixed site order, $\mathcal{O}(\chi^3 d^3)$, and becomes extremely large, especially when $N_{\rm s}$ is large.
Therefore, reducing the cost is essential for achieving an efficient site-order optimization.
Below, we discuss methods to reduce the computational cost, and the resulting costs are summarized in Table\ \ref{tab:comp_cost}.

A simple method to reduce computational overhead is a recursive strategy that caches and reuses intermediate tensors generated during successive SVD steps.
This approach reduces the cost to $\mathcal{O}(N_{\rm s}! \chi^3 e^{d}d^{N_{\rm s}})$.
(See Appendix\ \ref{app:store_reuse} for details of this derivation.)

Furthermore, one can adopt another approach based on the fact that the entanglement entropy between two subsystems remains invariant under arbitrary rearrangements of sites within each subsystem.
This property of entanglement enables us to obtain the entanglement entropies on the auxiliary bonds for $N_{\rm s}!$ candidates of site rearrangements by computing the entanglement entropies only for a minimal required set of bipartitions of sites and combining the values according to each candidate. (See Appendix\ \ref{app:lookup}.)
This approach to use a lookup table of these entanglement values reduces the computational cost to $\mathcal{O}(2^{N_s}\chi^{3}d^{N_s+\lfloor N_s/2\rfloor})$.
In addition, since this method does not rely on a recursive bookkeeping of decomposition paths, it avoids the growth of memory usage associated with recursive procedures.
A similar approach was employed also for a structural search algorithm in the context of Tensor Cross Interpolation(TCI)~\cite{TreeTCI}.

In the techniques discussed above, no truncation for the bond dimension is assumed at the SVDs.
If we introduce the truncation with the upper bound $\chi$, the computational cost of the former method to store and reuse intermediate tensors is further reduced to $\mathcal{O}(N_{\rm s}! \chi^3 e^{d}d^{2})$, while that of the latter method to utilize the lookup table of entanglement becomes $\mathcal{O}(2^{N_s}\chi^{3}d^{N_s})$.
(See Appendices\ \ref{app:store_reuse} and \ref{app:lookup}, respectively.)
Such processes with truncation can be adopted as approximations.
The effect of truncation, however, differs qualitatively between the two approaches.  
In the sequential-SVD scheme of Appendix~\ref{app:store_reuse}, a truncation is applied repeatedly during 
the process, and the resulting approximation errors may propagate to subsequent SVDs.  
In contrast, in the lookup-table strategy of Appendix~\ref{app:lookup}, each bipartition is treated 
independently, and the partial SVD with bond dimension~$\chi$ is not fed back into other SVDs.  
Therefore, no error propagation occurs by construction, making the truncation in the lookup-table method 
a more robust approximation.

Since the process of SVDs to evaluate the entanglement entropies for $N_{\rm s}!$ site orders involves lots of independent subprocesses, employing a parallel computation must also be effective to reduce the cost.
The details of the derivation of the costs reduced by the parallelization are presented in Appendix\ \ref{app:parallel}.

In practical calculations, performing the site order optimization with a small bond dimension is also a promising strategy.
As we will see in Sec.\ \ref{subsec:results}, the site-order optimization performed even with a small $\chi$ can identify a good site order that yields a significantly low variational energy.
Therefore, one can expect that such a site-order optimization with a small $\chi$ can achieve a remarkable accuracy improvement with reasonably low computational cost.

\section{Numerical results}\label{sec:NumRes}
\subsection{Model and calculation}\label{subsec:model}

\begin{figure}
\begin{center}
\includegraphics[width = 60 mm]{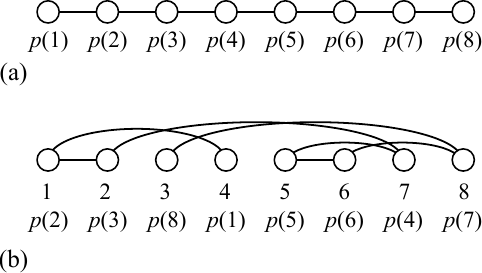}
\caption{
(a) Spin model in a 1D lattice with random site permutation.
(b) The same model as (a) but in a 1D lattice where the sites are arranged in ascending order.
An example for the case of an eight-site system with $\{ p(1), p(2), ..., p(8)\} = \{ 4, 1, 2, 7, 5, 6, 8, 3 \}$ is shown.
}
\label{fig:model}
\end{center}
\end{figure}

In order to evaluate the performance of the algorithm described in the previous section, we have applied it to the spin-1/2 Heisenberg model in the 1D lattice with random site order, as a benchmark model.
The model Hamiltonian is given by
\begin{eqnarray}
\mathcal{H} = J \sum_{i=1}^{N-1} {\bm S}_{p(i)} \cdot {\bm S}_{p(i+1)},
\label{eq:Ham}
\end{eqnarray}
where $N$ is the number of spins and $J$ is the exchange coupling constant.
We set $J=1$ in the following.
$\{ p(i)\}$ is a random permutation of the site index $\{ 1, 2, ..., N\}$.
If one arranges the sites in the order of $\{ p(i)\}$, the model (\ref{eq:Ham}) is nothing but the spin-1/2 nearest-neighbor Heisenberg chain under open boundary conditions [Fig.\ \ref{fig:model}(a)].
On the other hand, if the sites are arranged in the ascending order $\{ 1, 2, ..., N\}$, the model is a spin chain where the spins are coupled randomly but in a 1D single-stroke configuration [Fig.\ \ref{fig:model}(b)].
This model appears simple, but in practice, is extremely hard to tackle by the DMRG method without the site order optimization.
Namely, if one uses the MPS with an unoptimized site order (say, the ascending one), the bond dimension required to achieve an accurate calculation beyond a certain precision increases exponentially with the system size $N$.
We note that the optimal site order for this model is naturally $\{ p(i)\}$.

We have performed the calculations of the site-order optimization algorithm for the model (\ref{eq:Ham}) with the random site permutation $\{ p(i) \}$.
The site order optimization is specified by the range of the local site rearrangement, $N_{\rm s}$, and the bond dimension used during the optimization, $\chi_{\rm opt}$.
The optimization has started from an initial MPS where the sites are arranged in ascending order $\{ 1, 2, ..., N\}$ and yielded an MPS with an optimized site order $\{ r(i) \}$.
Then, we have performed the standard DMRG calculation with bond dimension $\chi_{\rm D}$ using the optimized site order $\{ r(i) \}$.
Below, we examine the accuracy of the thus-obtained variational energy and the optimized site order $\{ r(i) \}$ to evaluate the performance of the site-order optimization algorithm.

In our calculation, we have treated the system with $N=16, 32$ spins, and the number of random samples is $\mathcal{N}=100$. 
The site-order optimization was performed for $N_{\rm s} = 2 - 6$ and $\chi_{\rm opt} = 16, 32, 48$ ($16, 32$) for systems with $N = 16$ ($32$) spins.
The DMRG calculation using the optimized site order $\{ r(i) \}$ has been performed for 100 sweeps with $\chi_{\rm D}=16, 32, 48$, and the lowest energy obtained throughout all sweeps has been adopted as the variational energy.
In the following, we present only the results for $\chi_{\rm D}=48$, since the results for $\chi_{\rm D}=16$ and $32$ exhibit qualitatively identical behavior.

\subsection{Results}\label{subsec:results}

First, we examine the relative error in the variational energy,
\begin{eqnarray}
\delta e(\nu) = \frac{E_0(\nu) - E_{\rm exact}}{|E_{\rm exact}|},
\label{eq:energy_err}
\end{eqnarray}
where $E_0(\nu)$ is the variational energy obtained by the standard DMRG using the optimized site order $\{ r(i)\}$ for a random sample $\nu \in [1, \mathcal{N}]$.
$E_{\rm exact}$ is the numerically exact ground-state energy obtained by the exact diagonalization.

\begin{figure}
\begin{center}
\includegraphics[width = 70 mm]{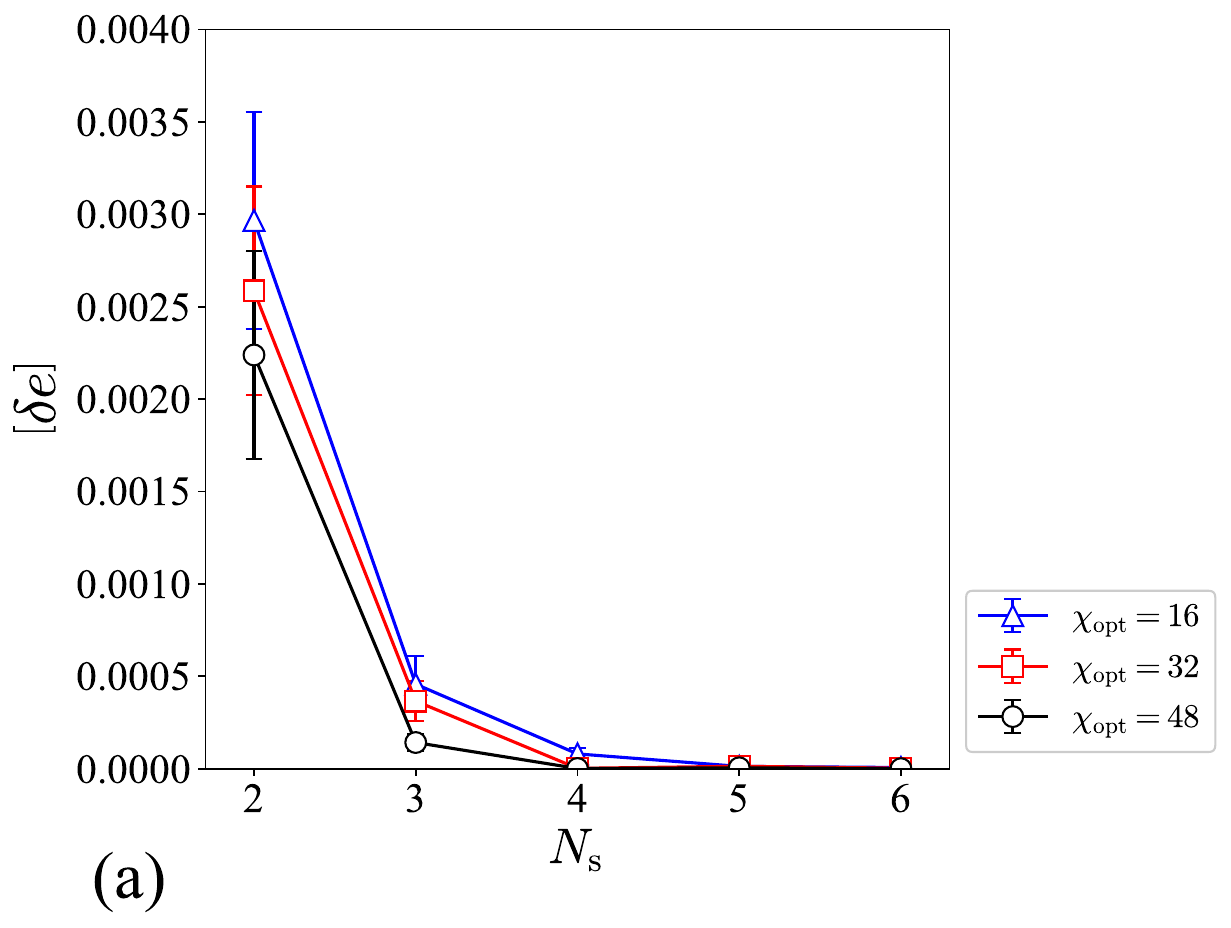}
\includegraphics[width = 70 mm]{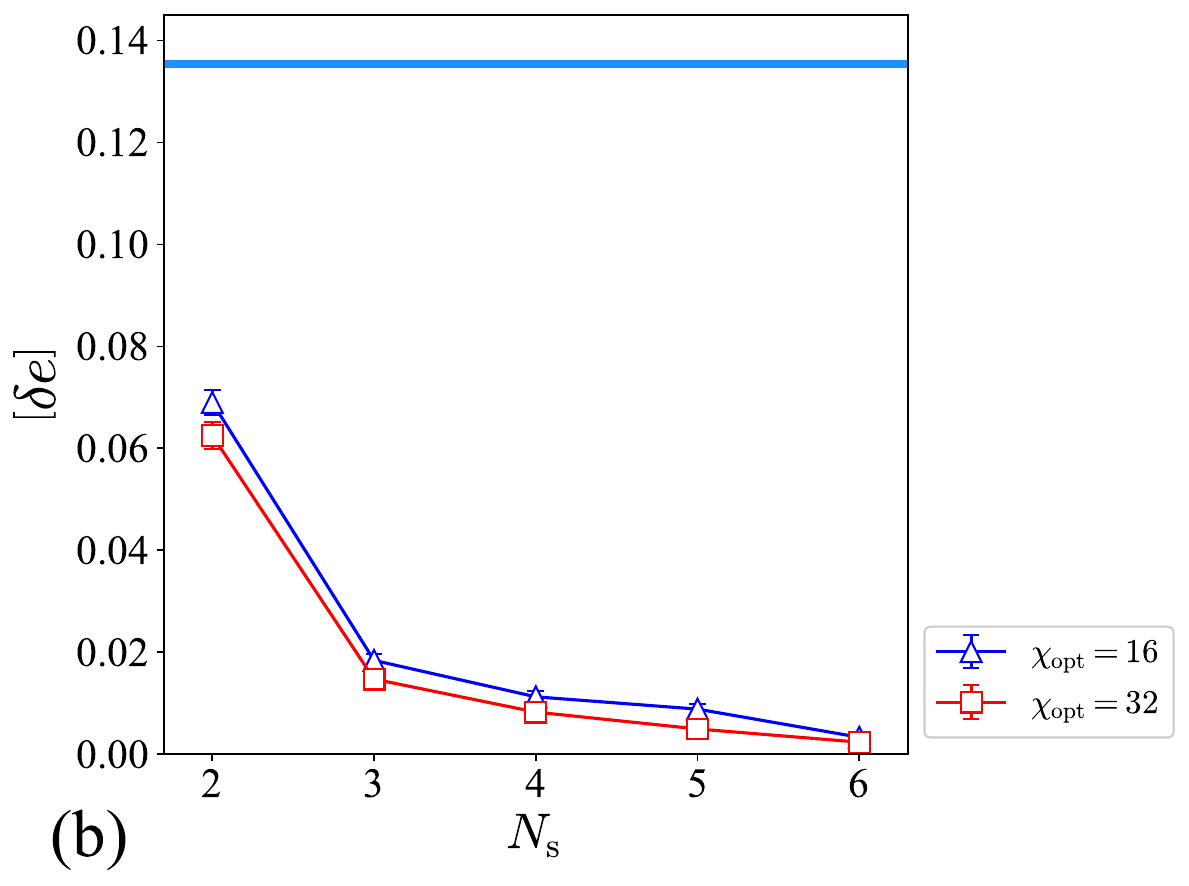}
\caption{
Random average of the relative error in variational energy, $[\delta e]$, for (a) $N=16$ and (b) $N=32$.
The variational energy is computed by the standard DMRG with $\chi_{\rm D}=48$ using the site order $\{ r(i)\}$ determined by the site order optimization with $N_{\rm s}$ and $\chi_{\rm opt}$.
In (a) [(b)], the data for $\chi_{\rm opt}=16, 32, 48$ ($16,32$) are plotted as a function of $N_{\rm s}$.
Horizontal blue line in (b) represents the value of $[\delta e] \sim 0.135$ obtained for $N=32$ and $\chi_{\rm D}=48$ without the site order optimization, i.e., $[ \delta e ]$ obtained using the ascending site order $\{ r(i)=i\}$.
}
\label{fig:rltvgeerr}
\end{center}
\end{figure}

Figure\ \ref{fig:rltvgeerr} shows the average of relative error $\delta e(\nu)$ over random samples, $[\delta e]$, obtained by the standard DMRG with $\chi_{\rm D}=48$ using the site order $\{ r(i)\}$ determined by the site order optimization with $(N_{\rm s}, \chi_{\rm opt})$.
$[ O ]$ denotes the random average of the quantity $O$ taken over $\mathcal{N}$ samples.
We also indicate in Fig.\ \ref{fig:rltvgeerr} (b) the value of $[ \delta e]$, $[ \delta e] \sim 0.135(7)$, for $N=32$ and $\chi_{\rm D}=48$ obtained without the site order optimization, that is, obtained from the standard DMRG using the ascending site order.
[For Fig.\ \ref{fig:rltvgeerr} (a), the value $[\delta e] \sim 0.022(5)$ obtained without the site order optimization for $N=16$ and $\chi_{\rm D}=48$ is not shown as it is too much larger than the data with the site order optimization.]
Figure\ \ref{fig:rltvgeerr} demonstrates that the site order optimization achieves significant reduction of the relative error $[\delta e]$ from the values without the optimization.
$[\delta e]$ becomes smaller when the range of site rearrangement is larger, as expected.
In particular, increasing the range from $N_{\rm s}=2$ to 3 realizes a remarkable reduction of $[\delta e]$, by 65\% to 94\% in the cases we tested.
As for the $\chi_{\rm opt}$ dependence, $[\delta e]$ also decreases as $\chi_{\rm opt}$ increases.
However, the effect of $\chi_{\rm opt}$ is minor compared to the effect of increasing $N_{\rm s}$ from 2 to 3.

\begin{figure}
\begin{center}
\includegraphics[width = 70 mm]{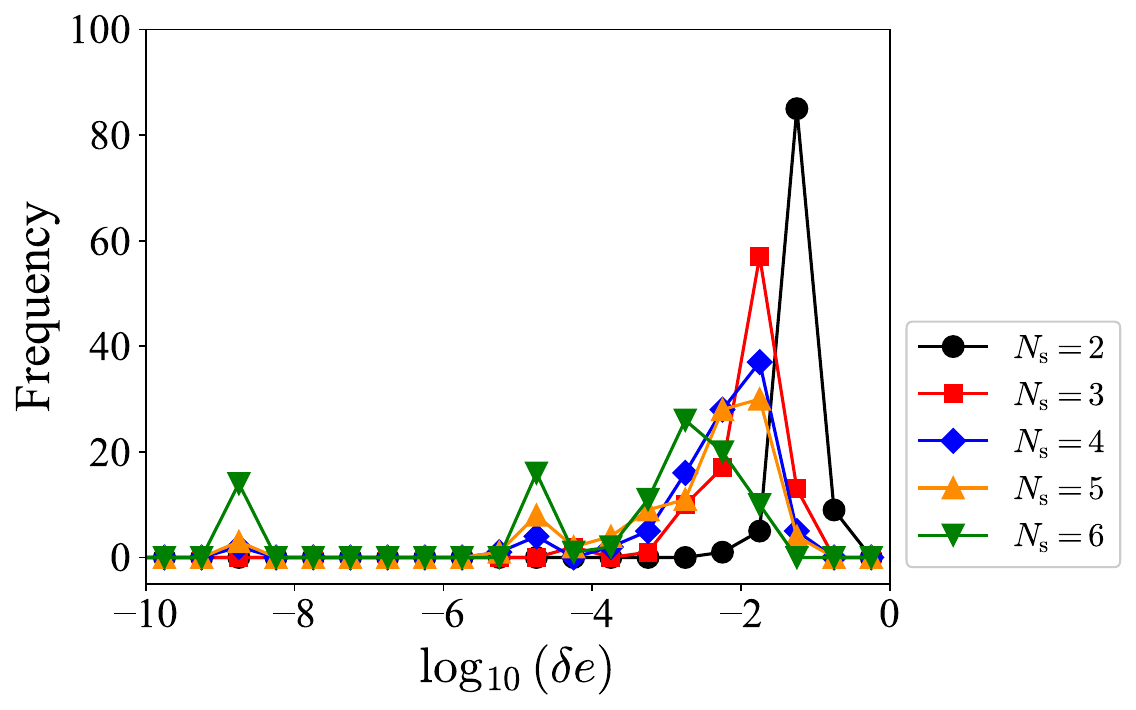}
\caption{
Frequency of histogram of $\log_{10}(\delta e)$ for the relative error $\delta e(\nu)$ obtained from the standard DMRG calculations with $\chi_{\rm D}=48$ using the site order $\{ r(i)\}$ determined by the optimization with $\chi_{\rm opt}=16$ and $N_{\rm s}=2$-$6$.
The system size is $N=32$, and the total number of random samples is $\mathcal{N}=100$.
The bin size is $0.5$.
}
\label{fig:hist_rltvgeerr}
\end{center}
\end{figure}

Figure\ \ref{fig:hist_rltvgeerr} presents the histogram of the relative error $\delta e(\nu)$ obtained from the DMRG calculations with $\chi_{\rm D}=48$ using the site order determined by the optimization with $N_{\rm s}=2$ - $6$ and $\chi_{\rm opt} = 16$ for 
$N=32$ spins.
The histogram is created for the logarithm of the relative error, with the bin size of $0.5$.
The frequency distribution for $N_{\rm s}=2$ exhibits a high peak at large errors around $\log_{10}(\delta e) \sim -1$.
As $N_{\rm s}$ increases, the peak shifts toward smaller errors and the peak height decreases.
The result also indicates that as $N_{\rm s}$ increases, the optimized site order can yield a smaller error in the variational energy, which is consistent with the behavior of $[\delta e]$.

Next, we investigate the optimized site order itself.
Specifically, we examine how much the distance of exchange interactions has been reduced by the optimization.
Consider the MPS with the optimized site order $\{ r(i)\}$.
Then, the distance between the two sites $p(i)$ and $p(i+1)$, that are connected by the $i$th exchange interaction, is given by
\begin{eqnarray}
\mathcal{D}_i = |r^{-1}[p(i)] - r^{-1}[p(i+1)]|,
\label{eq:distance_Di}
\end{eqnarray}
where $r^{-1}(j)=i$ for $r(i)=j$.
We investigate the average of the distance $\mathcal{D}_i$ over all the exchange interactions,
\begin{eqnarray}
\mathcal{D}(\nu) = \frac{1}{N-1} \sum_{i=1}^{N-1} \mathcal{D}_i,
\label{eq:DistJav}
\end{eqnarray}
for each random sample.
We note that since the distance $\mathcal{D}_i$ [Eq.\ (\ref{eq:distance_Di})] is equivalent to the number of auxiliary bonds that the interaction crosses, its average [Eq.\ (\ref{eq:DistJav})] is expected to be strongly correlated with the strength of the entanglement on the auxiliary bonds in MPS. (See the discussion regarding Fig.\ \ref{fig:EEstructure} in Sec.\ \ref{subsec:siteorder}.)
If the optimization algorithm achieves the optimal site order, i.e., $r(i)=p(i)$, $\mathcal{D}(\nu)$ takes the minimum value of $\mathcal{D}(\nu)=1$.

\begin{figure}
\begin{center}
\includegraphics[width = 70 mm]{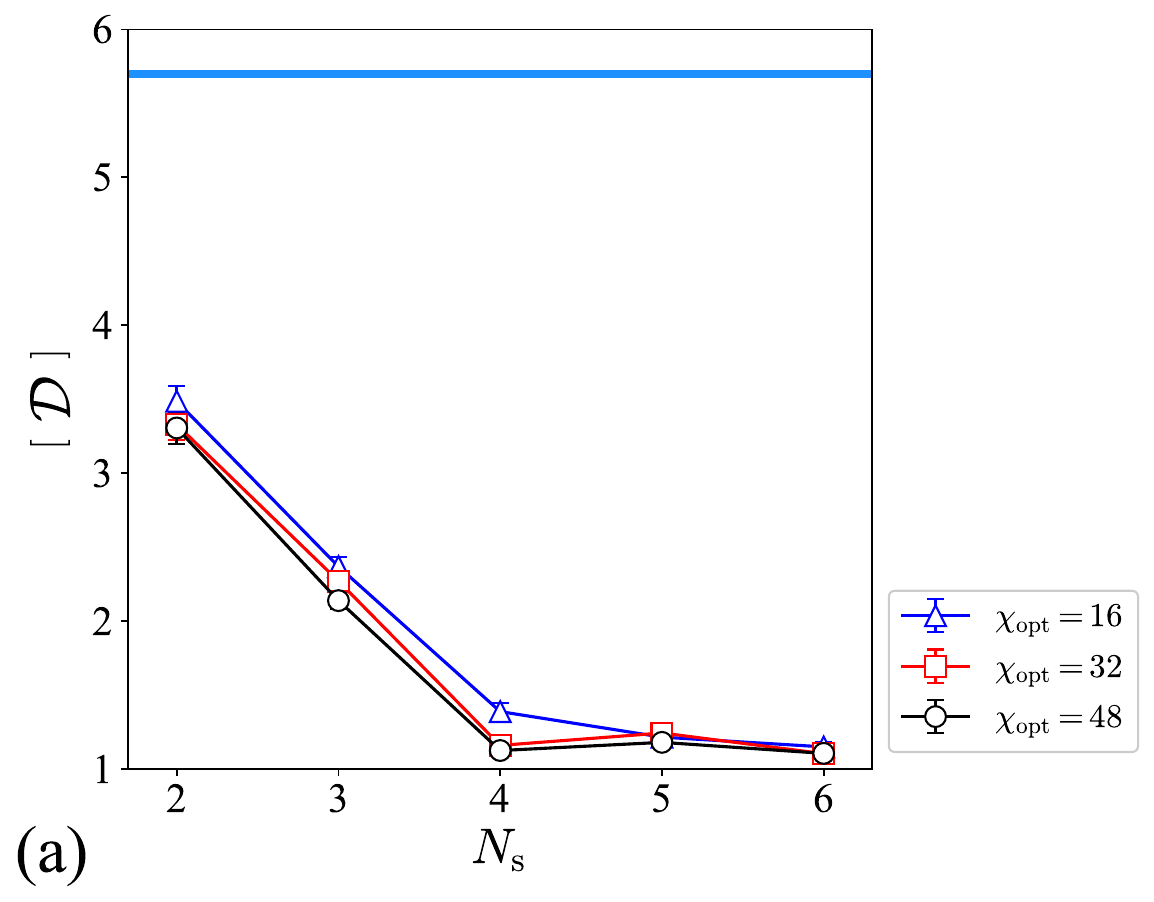}
\includegraphics[width = 70 mm]{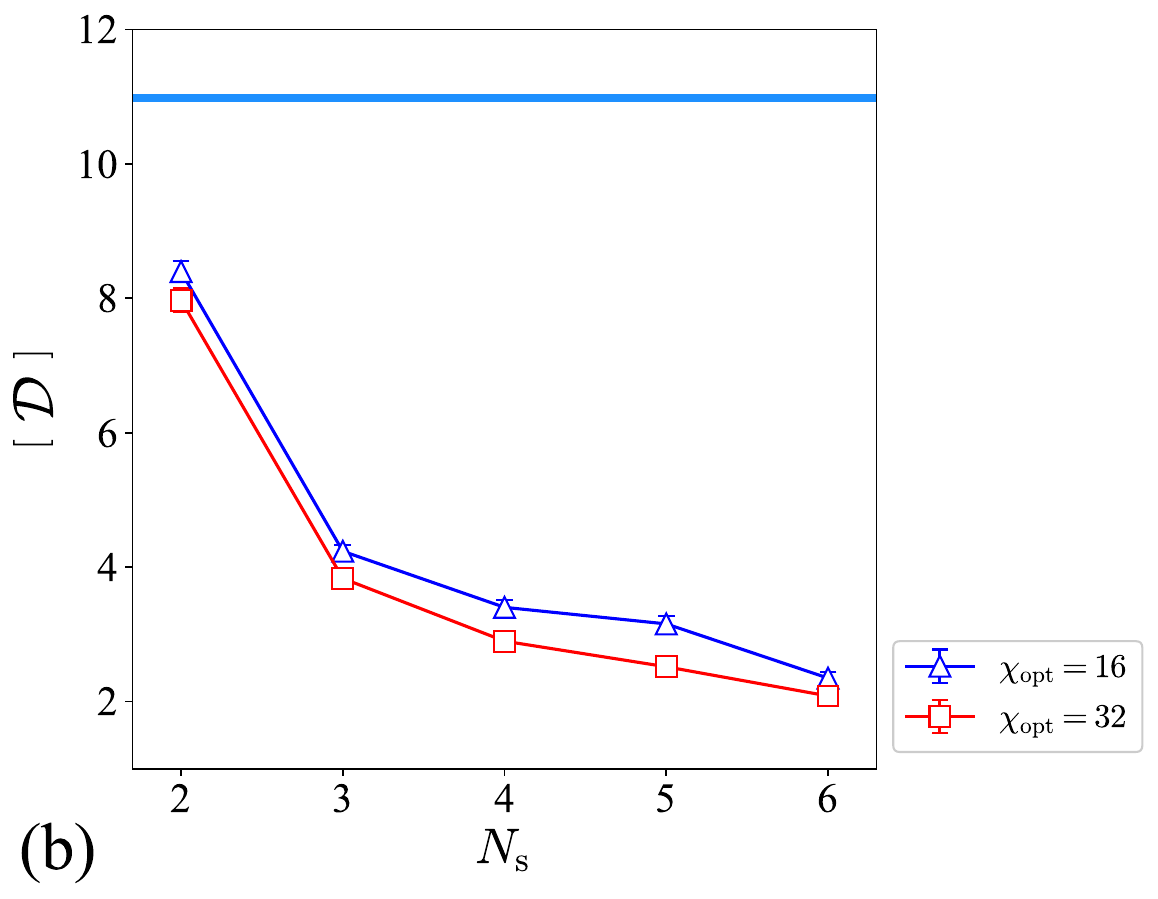}
\caption{
Random average of the interaction distance, $[\mathcal{D}]$, in the MPS with the optimized site order $\{ r(i)\}$ for (a) $N=16$ and (b) $N=32$.
In (a) [(b)], the data for $\chi_{\rm opt}=16, 32, 48$ ($16,32$) are plotted as a function of $N_{\rm s}$.
Horizontal blue line represents the value of $[\mathcal{D}]$ obtained without the site order optimization, i.e., $[\mathcal{D}]$ for the ascending site order $\{ r(i)=i\}$.
}
\label{fig:DistJ}
\end{center}
\end{figure}

Figure\ \ref{fig:DistJ} shows the result of the random average of $\mathcal{D}(\nu)$, $[\mathcal{D}]$, obtained by the site-order optimization with $\chi_{\rm opt}=16, 32, 48$ ($16, 32$) for the system with $N=16$ ($32$) spins.
For comparison, we also present the values of $[\mathcal{D}]$ obtained without the site order optimization, i.e., $[\mathcal{D}]$ for the ascending order $\{ r(i)=i\}$.
The data exhibit a similar dependence on $N_{\rm s}$ and $\chi_{\rm opt}$ to that observed for the relative error in the energy, $[\delta e]$;
Namely, $[\mathcal{D}]$ decreases significantly as $N_{\rm s}$ increases from $2$ to $3$, then decreases gradually with the further increase in $N_{\rm s}$.
$[\mathcal{D}]$ also decreases with increasing $\chi_{\rm opt}$, but this effect is secondary compared to that of increasing $N_{\rm s}$ from 2 to 3.
These results suggest that the accuracy improvement in the variational energy is indeed achieved through the improvement of the site order.

To conclude this section, we discuss the computational cost.
As discussed in Sec.\ \ref{subsec:comp_cost}, the site-order optimization algorithm requires a computational cost whose dependence on $N_{\rm s}$ and $d$ is more severe compared to the cost of the standard DMRG with a fixed site order.
Nevertheless, it is noteworthy that the site-order optimization with a small $\chi_{\rm opt}$ can achieve significant accuracy improvement, as we have seen above.
For instance, suppose that the site-order optimization is implemented using the method to use the table of entanglement [Method (b) in Table.\ \ref{tab:comp_cost}] without truncation.
The computational cost $\mathcal{O}(2^{N_s}\chi^{3}d^{N_s+\lfloor N_s/2\rfloor})$ with $\chi_{\rm opt}=16$ and $N_{\rm s}=3$ is smaller than or comparable to that of the standard DMRG, $\mathcal{O}(\chi_{\rm D}^3 d^3)$ with $\chi_{\rm D}=48$.
With such a small cost, the DMRG with $\chi_{\rm D}=48$ using the site order {\it optimized} with those $N_{\rm s}$ and $\chi_{\rm opt}$ has yielded the relative error in the variational energy, $[\delta e] \sim 0.018(1)$ for $N=32$, which is much smaller than that obtained by the DMRG with $\chi_{\rm D}=48$ without the site order optimization, $[\delta e] \sim 0.135(7)$.
The result demonstrates the effectiveness of the site-order optimization as a preprosessing treatment.

\section{Conclusion}\label{sec:conclusion}

We have investigated the effect of site order optimization on the accuracy of DMRG calculations.
We have focused on an algorithm that searches for the optimal site order by locally rearranging the sites\cite{LiRYS2022} and improved the algorithm by expanding the range of site rearrangement.
Then, we have applied the algorithm to a 1D Heisenberg model with randomly permuted site order.
The results demonstrate that the increase in the range of local site rearrangement, even a small increase from 2 to 3, can significantly improve the accuracy of the DMRG calculations.
Notably, this enhancement remains substantial even if the site-order optimization is performed with a small bond dimension $\chi_{\rm opt}$.

The results of the present study indicate that the site-order optimization algorithm offers an effective preprocessing method for MPS-based approaches, including the DMRG calculations.
The integration of the algorithm with existing low-cost site-order optimization schemes~\cite{ChanHG2002,LegezaS2003,MoritzHR2004,LegezaVPD2015} as a further preprocessing step also holds great potential for enhancing performance.

An important future development is applying the site-order optimization algorithm with multiple-site rearrangement to different systems and problems.
The application to fermionic systems is particularly relevant for quantum chemistry and other fields.
This implementation can be achieved simply by adopting a DMRG solver for fermionic systems in the part of the algorithm to optimize isometries (Process 1 in Table\ \ref{tab:algorithm}).
Application beyond quantum many-body systems, such as machine learning\cite{MPS_ML} and TCI\cite{Nunezetal2025}, must also be intriguing.

Accelerating the computations is also a critical issue.
The process requiring the highest cost is the SVDs to evaluate the entanglement entropies for $N_{\rm s}!$ candidates of possible site order.
Parallelization is a simple but powerful approach for accelerating the calculation.
Monte Carlo-like improvements, such as stochastically selecting the site order from a small number of randomly chosen candidates rather than all $N_{\rm s}!$ ones, are also expected to be promising.

The structural optimization of tensor networks through local network rearrangement has also been applied to tree-tensor networks (TTNs)\cite{Larsson2019,HikiharaUOHN2023,HikiharaUOHN2025,WatanabeMHU2026}.
The improvement of the algorithm by expanding the range of network rearrangement can be straightforwardly extended to the TTN approach.
However, for TTNs, the computational cost of the algorithm with rearrangement of $N_{\rm s}$ isometries becomes enormously large compared to that for MPS, under the usual situation where $\chi \gg d$.
To implement the improvement on TTN, it is necessary to find a solution to suppress the cost, which is left for future research.

Beyond the structural optimization, our results also provide important insights for augmented DMRG methods~\cite{Qian_2023,PhysRevLett.133.190402,huang2025augmentingdensitymatrixrenormalization}.
As increasing $N_s$ in the site order optimization improves accuracy, we expect that enlarging the size of disentanglers in augmented DMRG could further enhance numerical precision, offering a promising direction for the future of tensor network simulations.

\section*{Acknowledgments}
We acknowledge the use of ITensors and ITensorMPS libraries\cite{iTensor,iTensor03} to implement the algorithm.
This work is partially supported by KAKENHI Grant Numbers JP24K06881, JP22H01171, JP21H04446, JP21H05182, JP21H05191, and JP25KJ1773 from JSPS of Japan.
We also acknowledge support from MEXT Q-LEAP Grant No.
JPMXS0120319794, and from JST COI-NEXT No. JPMJPF2014, ASPIRE No. JPMJAP2319, and CREST No. JPMJCR24I1 and JPMJCR24I3.
R.W. thanks the discussion with H. Shinaoka.
H.U. was supported by the COE research grant in computational science from Hyogo
Prefecture and Kobe City through Foundation for Computational Science.

\section*{Data Availability}
The code used to generate the results in this work is available at \cite{FlexibleDMRG}.
The data supporting the findings of this study are not publicly available due to practical limitations related to data preparation and storage.
However, the data can be obtained from the authors upon reasonable request.

\appendix

\section{
Computational cost of sequential SVDs for site rearrangement
}\label{app:store_reuse}

In this appendix, we derive the computational cost of the sequential
singular value decompositions (SVDs) used to evaluate the entanglement entropies for $N_{\rm s}!$ possible site orders in the focused region of MPS.
We assume that no truncation of singular values is applied and that all
intermediate tensors (singular values and singular vectors) obtained at
each step are stored and reused in the subsequent steps.

We begin with the $(N_{\rm s}+2)$-leg tensor obtained by contracting
$N_{\rm s}$ isometries and a singular-value matrix in the focused region
[see Fig.~\ref{fig:algorithm}].
Its index structure is
\begin{equation}
    \{\alpha_{\rm L}\},
    \{\sigma_{f},\dots,\sigma_{f+N_{\rm s}-1}\},
    \{\alpha_{\rm R}\}~,
\end{equation}
where the auxiliary indices $\alpha_{\rm L}$ and $\alpha_{\rm R}$ have
dimension $\chi$, and each physical index $\sigma_j$ has dimension $d$.

The sequential SVD proceeds by extracting one physical index at a time
and attaching it to the left block.
At the first SVD step, we select one of the $N_{\rm s}$ physical indices, say $\sigma_{i_1}$, and reshape the tensor into a matrix by grouping the
indices as follows:
\begin{itemize}
\item left block (columns): $(\alpha_{\rm L}, \sigma_{i_1})$, of dimension $\chi d$,
\item right block (rows): the remaining physical indices and $\alpha_{\rm R}$, of dimension $\chi d^{N_{\rm s}-1}$.
\end{itemize}
Applying an SVD yields a left isometry and a right tensor that
contains the remaining $N_{\rm s}-1$ physical indices.
These singular vectors and singular values are stored and reused for all
candidate permutations in subsequent steps.


At the second SVD step, for each of the $N_{\rm s}-1$ remaining physical indices, we again form an SVD matrix by attaching the chosen physical index to the updated left block. Importantly, although no truncation is assumed, the SVD is performed on a rectangular matrix, and hence the effective dimension of the left block is bounded by the maximal rank allowed by the right block.

More generally, after the $(i-1)$-th SVD step, the number of physical indices absorbed into the left block is $i-1$.
In the absence of rank constraints, the left block would have dimension $\chi d^{i-1}$.
However, since the SVD at the $(i-1)$-th step is performed on a rectangular matrix whose right block has dimension $\chi d^{N_{\rm s}-i+1}$, the effective bond dimension is bounded as
\begin{equation}
\rho_{i-1}
=
\chi d^{\min(i-1,N_{\rm s}-i+1)}.
\end{equation}

At the $i$-th SVD step, one selects one of the $(N_{\rm s}-i+1)$ remaining physical indices and reshapes the tensor into an $L_i\times M_i$ matrix with
\begin{align}
L_i &= \rho_{i-1} d
= \chi d^{\min(i,N_{\rm s}-i+2)},
\label{eq:Li_def}\\
M_i &= \chi d^{N_{\rm s}-i}.
\label{eq:Mi_def}
\end{align}

Since the computational cost of an SVD for an $L\times M$ matrix is
$O(\max(L,M)[\min(L,M)]^{2})$, the cost of the $i$-th SVD scales as $ O\!\left(\chi^3 d^{N_{\rm s}}d^{E_i}\right)$ with 
\begin{equation}
E_i=
\begin{cases}
i, & 1\le i\le \lfloor N_{\rm s}/2\rfloor,\\
N_{\rm s}-i, & \lfloor N_{\rm s}/2\rfloor < i\le \lfloor (N_{\rm s}+2)/2\rfloor,\\
2N_{\rm s}-3i+2, & \lfloor (N_{\rm s}+2)/2\rfloor < i\le N_{\rm s}-1.
\end{cases}
\end{equation}

Taking into account the number of ordered choices of extracted physical indices,
\begin{equation}
\prod_{k=1}^{i}(N_{\rm s}-k+1)=\frac{N_{\rm s}!}{(N_{\rm s}-i)!},
\end{equation}
the total cost associated with the $i$-th SVD step scales as
\begin{equation}
T_i
=
\frac{N_{\rm s}!}{(N_{\rm s}-i)!}
\chi^3 d^{N_{\rm s}}d^{E_i},
\end{equation}
where constant prefactors independent of $N_{\rm s}$, $\chi$, and $d$ have been omitted.

The total computational cost of the sequential SVD procedure is thus given by
\begin{align}
T_{\rm tot}
&=
\chi^3 N_{\rm s}! d^{N_{\rm s}}
\sum_{i=1}^{N_{\rm s}-1} \frac{d^{E_i}}{(N_{\rm s}-i)!}
\end{align}

To bound the sum in the expression for $T_{\rm tot}$, we introduce $j=N_{\rm s}-i$ and rewrite
\begin{equation}
T_{\rm tot}
=
\chi^3 N_{\rm s}! d^{N_{\rm s}}
\sum_{j=1}^{N_{\rm s}-1}\frac{d^{\widetilde E_j}}{j!},
\qquad
\widetilde E_j:=E_{N_{\rm s}-j}.
\label{eq:cost_sum}
\end{equation}

For $d\ge 1$, we have $\widetilde{E}_j\le j$, and therefore
\( d^{\widetilde{E}_j} \le d^j \).
Using $\sum_{j=1}^\infty d^j/j! = e^d - 1$, the sum in
Eq.~\eqref{eq:cost_sum} obeys the uniform bound
\begin{equation}
\sum_{j=1}^{N_{\rm s}-1}
  \frac{d^{\widetilde{E}_j}}{j!}
\le
e^{d}-1
\end{equation}
for $N_{\rm s}\ge 2$.
Consequently, the total SVD cost obeys the upper bound
\begin{equation}
T_{\rm tot}
\le
(e^{d}-1) \chi^3 N_{\rm s}! d^{N_{\rm s}} \sim \mathcal{O}(N_{\rm s}! \chi^3 e^{d} d^{N_{\rm s}}),
\end{equation}
which holds for both even and odd $N_{\rm s}$.

\paragraph*{Remark on truncations.}
If one wishes to enlarge the range of site rearrangement, $N_{\rm s}$, it is natural to introduce truncations in sequential SVDs.
When a specific truncation with the bond dimension $\chi$ is introduced
in each SVD, the computational cost is modified.
In this case the $i$-th SVD cost becomes
\begin{equation}
T^{({\rm trunc})}_i
= \prod_{k=1}^{i} (N_{\rm s}-k+1)
  \chi^3 d^{N_{\rm s}-i+2},
\qquad i=1,\dots,N_{\rm s}-1 ,
\end{equation}
which reduces to
\begin{equation}
T^{({\rm trunc})}_i
= \frac{N_{\rm s}!}{(N_{\rm s}-i)!}
  \chi^3 d^{N_{\rm s}-i+2}.
\end{equation}
The total cost is then becomes
\begin{equation}
T^{({\rm trunc})}_{\rm tot}
= \chi^3 N_{\rm s}! 
  \sum_{i=1}^{N_{\rm s}-1} 
    \frac{d^{N_{\rm s}-i+2}}{(N_{\rm s}-i)!}
= \chi^3 N_{\rm s}! d^{2}
  \sum_{j=1}^{N_{\rm s}-1} \frac{d^{j}}{j!},
\end{equation}
where we have again introduced $j = N_{\rm s}-i$.
For $d\ge 1$, the truncated sum satisfies
\(
\sum_{j=1}^{N_{\rm s}-1} d^{j}/j!
 \le e^{d}-1,
\)
so that the total SVD cost with truncation obeys
\begin{equation}
T^{({\rm trunc})}_{\rm tot}
\le
(e^{d}-1) \chi^3 N_{\rm s}!  d^{2}
\sim \mathcal{O}(N_{\rm s}! \chi^3 e^{d} d^{2}).
\end{equation}


\section{Reducing the number of SVD calls by constructing a lookup table}
\label{app:lookup}

In this appendix, we show that the number of SVD calls required to
evaluate the entanglement entropies for all $N_{\rm s}!$ candidate site
orders can be substantially reduced by constructing a minimal lookup table
of SVD results.
This reduction is based on the fact that the entanglement entropy
associated with a bipartition depends only on which physical sites
belong to the left and right subsystems, and remains invariant under arbitrary permutation of sites within each subsystem.

\subsection*{B.1 \; Entanglement entropy depends only on the bipartition set}

Consider the tensor representing the wavefunction in the focused region, whose index structure is
\[
    \{\alpha_{\rm L}\},
    \{\sigma_{f},\dots,\sigma_{f+N_{\rm s}-1}\},
    \{\alpha_{\rm R}\},
\]
where $\alpha_{\rm L}$ and $\alpha_{\rm R}$ are auxiliary indices of
dimension $\chi$, and each physical index $\sigma_{j}$ has dimension $d$.
For a given bipartition that sends $i$ physical indices to the left, we
can write
\[
\text{left block: }
(\alpha_{\rm L}, \sigma'_{1},\dots,\sigma'_{i}),
\]
\[
\text{right block: }
(\sigma'_{i+1},\dots,\sigma'_{N_{\rm s}},\alpha_{\rm R}),
\]
where $\{\sigma'_{1},\dots,\sigma'_{N_{\rm s}}\}$ is any reordering of
$\{\sigma_{f},\dots,\sigma_{f+N_{\rm s}-1}\}$.

Crucially, the singular values obtained from the SVD for this
bipartition, and thus the entanglement entropy, is invariant under
arbitrary permutations within each of the left and right subsets.
Therefore, for a fixed subset of sites of size $i$, we only need to
perform the SVD for one representative ordering that satisfies
\[
\sigma'_{1} < \sigma'_{2} < \dots < \sigma'_{i},
\qquad
\sigma'_{i+1} < \dots < \sigma'_{N_{\rm s}},
\]
and the entanglement entropy for all other orderings of the same subset
is identical.

Since choosing the representative subset of size $i$ amounts to choosing
$i$ sites from $N_{\rm s}$, the number of distinct bipartitions for which
an SVD must be computed is exactly
\(
\binom{N_{\rm s}}{i}.
\)
Thus the total number of required SVDs is
\[
\sum_{i=1}^{N_{\rm s}-1} \binom{N_{\rm s}}{i}=2^{N_{\rm s}}-2,
\]
which is exponentially smaller than $N_{\rm s}!$.

Once these SVDs are computed, the entanglement entropies for all
$N_{\rm s}!$ permutations can then be evaluated simply by referencing the
lookup table.
Thus the expensive SVD calls are performed only once per bipartition
subset.

\subsection*{B.2 \; Cost of building the lookup table}

Let $c_i$ denote the total SVD cost associated with bipartitions in which
$i$ physical indices belong to the left subsystem and $N_{\rm s}-i$ to the
right. For each $i=1,\dots,N_{\rm s}-1$, the number of distinct subsets of
size $i$ is $\binom{N_{\rm s}}{i}$.
When no truncation is applied, the cost of a single SVD corresponding for such a
bipartition scales as
\[
\chi^3 d^{N_{\rm s} + \min(i,N_{\rm s}-i)},
\]
because the matrix dimensions are
$\chi d^{i} \times \chi d^{N_{\rm s}-i}$ and the SVD cost is
$O(\max(L,M)[\min(L,M)]^2)$.
Therefore,
\begin{equation}
c_i
= \binom{N_{\rm s}}{i}
  \chi^3 d^{N_{\rm s} + \min(i,N_{\rm s}-i)},
\qquad
i=1,\dots,N_{\rm s}-1.
\end{equation}

Summing over all bipartition sizes $i$ gives the total cost of
building the lookup table,
\begin{equation}
c_{\mathrm{tot}}
= \sum_{i=1}^{N_{\rm s}-1} c_i
= \chi^3 d^{N_{\rm s}}
  \sum_{i=1}^{N_{\rm s}-1}
    \binom{N_{\rm s}}{i}
    d^{\min(i,N_{\rm s}-i)}.
\label{eq:ctot_min_form}
\end{equation}
This expression is explicitly symmetric under $i\leftrightarrow N_{\rm s}-i$.
Note that a strict upper bound can also be obtained by using
$\min(i,N_{\rm s}-i)\le \lfloor N_{\rm s}/2 \rfloor$ for all $i$ and
$\sum_{i=1}^{N_{\rm s}-1}\binom{N_{\rm s}}{i}=2^{N_{\rm s}}-2$.
These give
\begin{equation}
c_{\mathrm{tot}}
\le
\chi^{3} d^{N_{\rm s}}
\left(2^{N_{\rm s}}-2\right)
d^{\lfloor N_{\rm s}/2 \rfloor} \sim O\!\left(
2^{N_{\rm s}} \chi^{3} d^{N_{\rm s}+\lfloor N_{\rm s}/2 \rfloor}
\right).
\end{equation}

\paragraph*{Remark on partial SVD.}
To further scale the lookup-table construction for larger $N_{\rm s}$, one can replace the full SVD with a \textit{partial} (rank-$\chi$) SVD for each bipartition.
This reduces the effective matrix dimension on the ``longer'' side of
the bipartition from $d^{N_{\rm s}/2}$ to $\chi$ which lowers the
asymptotic cost of the largest SVD from
$\chi^{3} d^{N_{\rm s}+\lfloor N_{\rm s}/2\rfloor}$
to $\chi^3d^{N_s}$.
We thus get reduced total cost
\begin{equation}
    c_{\rm{tot}}^{\mathrm{(partial)}}
    \sim \mathcal{O}(2^{N_{\rm s}}\chi^{3} d^{N_{\rm s}})~,
\end{equation}
up to a prefactor that differs from the full-SVD case.
Such a partial SVD does not retain the exact singular values, but is often a practical option when the goal is to explore large $N_{\rm s}$ in
an approximate yet scalable fashion.

\subsection*{B.3 \; Use of the lookup table}

Once all SVDs listed in Eq.~\eqref{eq:ctot_min_form} have been computed,
the lookup table contains the singular-value spectra (and hence the
entanglement entropies) for every bipartition of the $N_{\rm s}$ physical
indices into a subset of size $i$ and its complement of size $N_{\rm s}-i$
for all $i = 1,\dots, N_{\rm s}-1$.
Since entanglement entropy is invariant under site reordering within each subsystem, one can evaluate entanglement entropies for all $N_{\rm s}!$ candidate site orders without performing redundant SVDs:
each bipartition encountered during the sequential reconstruction of the
MPS is mapped to one of the bipartitions stored in the lookup table, and
its entanglement entropy is read directly from the precomputed value.

Once the site order yielding the minimal total entanglement entropy has been
identified, we perform a single sequence of SVDs (with truncation) to
reconstruct the MPS according to the optimal site order.
This single reconstruction step completes the site-order update for the
current optimization cycle.


\section{Parallelization and  computational cost of site-order optimizations}
\label{app:parallel}

\begin{table}[t]
\caption{
Cost of parallel computation for the SVD process to evaluate the entanglement entropies for $N_{\rm s}!$ candidates of possible site order.
(a) Method to store and reuse intermediate tensors and (b) Method to use the table of entanglement.
``Truncation'' refers to the procedure in which a full SVD is performed at every decomposition 
step and then the bond dimension is truncated to $\chi$.
In contrast, ``Truncation*'' refers to the use of a partial SVD that directly computes only $\chi$-largest
singular values and vectors without generating the full spectrum.
}
\label{tab:comp_cost_parallel}
\begin{center}
\begin{tabular}{cc}
\hline
\hline
\multicolumn{2}{l}{(a) Method to store and reuse intermediate tensors} \\
\hline
No truncation  &  $\mathcal{O}(\chi^3 d^{N_{\rm s}+\lfloor N_s/2 \rfloor})$  \\
Truncation    &  $\mathcal{O}(\chi^3 d^{N_{\rm s}+1})$  \\
\hline
\hline
\multicolumn{2}{l}{(b) Method to use table of entanglement} \\
\hline
No truncation &  $\mathcal{O}(\chi^{3}d^{N_s+\lfloor N_s/2\rfloor})$  \\
Truncation*   &  $\mathcal{O}(\chi^{3}d^{N_s})$  \\
\hline
\hline
\end{tabular}
\end{center}
\end{table}

In this appendix, we consider an idealized scenario in which sufficient computational resources are available to allow for trivial parallelization over independent SVDs.
Such parallelization can be applied to either (i)~the direct decomposition scheme described in Appendix~A or (ii)~the lookup-table construction scheme described in Appendix~B.
The number of computer nodes required to fully parallelize these approaches scales with the range of the site rearrangement, $N_{\rm s}$, as $N_{\rm s}!$ and $2^{N_{\rm s}}$, respectively.
Under this assumption of abundant resources, the multiplicative prefactors associated with the number of independent SVDs do not contribute to the wall-clock time, and only the cost of the individual SVD operations remains relevant.
The resulting computational costs for each method, both with and without truncation, are summarized in Table~\ref{tab:comp_cost_parallel}.

\subsection{Direct decomposition scheme (Appendix A)}

We first consider the direct SVD-based decomposition scheme summarized
in Appendix~A.
At the $i$-th decomposition step, the number of independent SVDs
associated with all ordered choices of extracted physical sites is
\(
  \prod_{k=1}^{i}(N_{\rm s}-k+1),
\)
but under trivial parallelization these SVDs can be executed simultaneously.
Thus the wall-clock cost at the $i$-th step is simply the cost of one SVD.

From Appendix~A, the cost of a single non-truncated SVD at the $i$-th
step scales as 
\begin{equation}
  t_i
  = \chi^{3} d^{N_{\rm s}} d^{E_i}~.
\end{equation}
Therefore the total wall-clock cost becomes
\begin{eqnarray}
  t_{\mathrm{tot}}
  = \sum_{i=1}^{N_{\rm s}-1} t_i \sim \mathcal{O}(\chi^{3} d^{N_{\rm s}+\lfloor N_{\rm s}/2 \rfloor} )~,
\end{eqnarray}
which reduces to the closed forms already derived in Appendix~A. 

When the truncation scheme discussed in Sec.~\ref{subsec:comp_cost} -- namely, truncating the bond dimension to $\chi$ after each SVD -- is applied at each step, the computational cost of the SVD at the $i$-th step is
\begin{equation}
  t^{({\rm trunc})}_i
  = \chi^{3} d^{N_{\rm s}-i+2},
  \qquad i=1,\dots,N_{\rm s}-1.
\end{equation}
Hence the total cost is given by a geometric series:
\begin{align}
  t^{({\rm trunc})}_{\mathrm{tot}}
  &= \sum_{i=1}^{N_{\rm s}-1} t^{({\rm trunc})}_i
     \sim \mathcal{O}(\chi^{3} d^{N_s+1})~.
\end{align}

\subsection{Lookup-table construction scheme (Appendix B)}

Appendix~B shows that the number of distinct bipartitions of the
$N_{\rm s}$ physical sites is $2^{N_{\rm s}}-2$, and each bipartition
requires one SVD.
These SVDs are mutually independent, making the lookup-table
construction ideally suited to trivial parallelization.

From Appendix~B, the cost of the SVD associated with a bipartition of
size $i$ (i.e.\ $i$ sites on the left and $N_{\rm s}-i$ on the right) is
\begin{equation}
  t_i = \chi^{3}
        d^{N_{\rm s} + \min(i,N_{\rm s}-i)}.
\end{equation}
Under trivial parallelization, all SVDs are executed simultaneously.
Hence the wall-clock cost for building the entire lookup table is determined by the largest SVD:
\begin{equation}
  t_{\mathrm{LT}}
  = \max_{1\le i\le N_{\rm s}-1}
    \left[
      \chi^{3} d^{N_{\rm s}+\min(i,N_{\rm s}-i)}
    \right].
\end{equation}
The exponent $\min(i,N_{\rm s}-i)$ is maximized at the ``middle''
bipartitions:
\[
  \min(i,N_{\rm s}-i)
  =
  \begin{cases}
    M, & N_{\rm s} = 2M, \\
    M, & N_{\rm s} = 2M+1,
  \end{cases}
\]
so we obtain
\begin{equation}
    t_{\mathrm{LT}}
    = \chi^{3} d^{N_{\rm s} + \lfloor N_{\rm s}/2 \rfloor}~.
\end{equation}

By employing a partial SVD, as discussed in the remark in Appendix B.2, the wall-clock cost of building the lookup table scales as
\begin{equation}
    t^{\rm (partial)}_{\mathrm{LT}} = \chi^{3} d^{N_{\rm s}}.
\end{equation}

Once the optimal site order is identified from the lookup table, only a single sequential SVD reconstruction of the MPS remains to be performed.
This step may involve truncation, and its computational cost is given by
\begin{equation}
  O\!\left(
    \chi^{3} d^{3}
    \frac{d^{N_{\rm s}-1}-1}{d-1}
  \right).
\end{equation}
\\
Since this reconstruction cost scales as $O(\chi^3 d^{N_s+1})$, 
i.e., with the same exponential dependence on $N_s$ as the lookup-table construction,
it does not increase the overall computational complexity.


\begin{thebibliography}{40}%
\makeatletter
\providecommand \@ifxundefined [1]{%
 \@ifx{#1\undefined}
}%
\providecommand \@ifnum [1]{%
 \ifnum #1\expandafter \@firstoftwo
 \else \expandafter \@secondoftwo
 \fi
}%
\providecommand \@ifx [1]{%
 \ifx #1\expandafter \@firstoftwo
 \else \expandafter \@secondoftwo
 \fi
}%
\providecommand \natexlab [1]{#1}%
\providecommand \enquote  [1]{``#1''}%
\providecommand \bibnamefont  [1]{#1}%
\providecommand \bibfnamefont [1]{#1}%
\providecommand \citenamefont [1]{#1}%
\providecommand \href@noop [0]{\@secondoftwo}%
\providecommand \href [0]{\begingroup \@sanitize@url \@href}%
\providecommand \@href[1]{\@@startlink{#1}\@@href}%
\providecommand \@@href[1]{\endgroup#1\@@endlink}%
\providecommand \@sanitize@url [0]{\catcode `\\12\catcode `\$12\catcode
  `\&12\catcode `\#12\catcode `\^12\catcode `\_12\catcode `\%12\relax}%
\providecommand \@@startlink[1]{}%
\providecommand \@@endlink[0]{}%
\providecommand \url  [0]{\begingroup\@sanitize@url \@url }%
\providecommand \@url [1]{\endgroup\@href {#1}{\urlprefix }}%
\providecommand \urlprefix  [0]{URL }%
\providecommand \Eprint [0]{\href }%
\providecommand \doibase [0]{https://doi.org/}%
\providecommand \selectlanguage [0]{\@gobble}%
\providecommand \bibinfo  [0]{\@secondoftwo}%
\providecommand \bibfield  [0]{\@secondoftwo}%
\providecommand \translation [1]{[#1]}%
\providecommand \BibitemOpen [0]{}%
\providecommand \bibitemStop [0]{}%
\providecommand \bibitemNoStop [0]{.\EOS\space}%
\providecommand \EOS [0]{\spacefactor3000\relax}%
\providecommand \BibitemShut  [1]{\csname bibitem#1\endcsname}%
\let\auto@bib@innerbib\@empty
\bibitem [{\citenamefont {Schollw\"ock}(2011)}]{SCHOLLWOCK2011}%
  \BibitemOpen
  \bibfield  {author} {\bibinfo {author} {\bibfnamefont {U.}~\bibnamefont
  {Schollw\"ock}},\ }\bibfield  {title} {\bibinfo {title} {The density-matrix
  renormalization group in the age of matrix product states},\ }\href
  {https://doi.org/https://doi.org/10.1016/j.aop.2010.09.012} {\bibfield
  {journal} {\bibinfo  {journal} {Annals of Physics}\ }\textbf {\bibinfo
  {volume} {326}},\ \bibinfo {pages} {96} (\bibinfo {year} {2011})},\ \bibinfo
  {note} {january 2011 Special Issue}\BibitemShut {NoStop}%
\bibitem [{\citenamefont {Verstraete}\ \emph {et~al.}(2023)\citenamefont
  {Verstraete}, \citenamefont {Nishino}, \citenamefont {Schollw\"ock},
  \citenamefont {Ba\~nuls}, \citenamefont {Chan},\ and\ \citenamefont
  {Stoudenmire}}]{Review2023}%
  \BibitemOpen
  \bibfield  {author} {\bibinfo {author} {\bibfnamefont {F.}~\bibnamefont
  {Verstraete}}, \bibinfo {author} {\bibfnamefont {T.}~\bibnamefont {Nishino}},
  \bibinfo {author} {\bibfnamefont {U.}~\bibnamefont {Schollw\"ock}}, \bibinfo
  {author} {\bibfnamefont {M.~C.}\ \bibnamefont {Ba\~nuls}}, \bibinfo {author}
  {\bibfnamefont {G.~K.}\ \bibnamefont {Chan}},\ and\ \bibinfo {author}
  {\bibfnamefont {M.~E.}\ \bibnamefont {Stoudenmire}},\ }\bibfield  {title}
  {\bibinfo {title} {Density matrix renormalization group, 30 years on},\
  }\href {https://doi.org/10.1038/s42254-023-00572-5} {\bibfield  {journal}
  {\bibinfo  {journal} {Nature Reviews Physics}\ }\textbf {\bibinfo {volume}
  {5}},\ \bibinfo {pages} {273} (\bibinfo {year} {2023})}\BibitemShut {NoStop}%
\bibitem [{\citenamefont {White}(1992)}]{White1992}%
  \BibitemOpen
  \bibfield  {author} {\bibinfo {author} {\bibfnamefont {S.~R.}\ \bibnamefont
  {White}},\ }\bibfield  {title} {\bibinfo {title} {Density matrix formulation
  for quantum renormalization groups},\ }\href
  {https://doi.org/10.1103/PhysRevLett.69.2863} {\bibfield  {journal} {\bibinfo
   {journal} {Phys. Rev. Lett.}\ }\textbf {\bibinfo {volume} {69}},\ \bibinfo
  {pages} {2863} (\bibinfo {year} {1992})}\BibitemShut {NoStop}%
\bibitem [{\citenamefont {White}(1993)}]{White1993}%
  \BibitemOpen
  \bibfield  {author} {\bibinfo {author} {\bibfnamefont {S.~R.}\ \bibnamefont
  {White}},\ }\bibfield  {title} {\bibinfo {title} {Density-matrix algorithms
  for quantum renormalization groups},\ }\href
  {https://doi.org/10.1103/PhysRevB.48.10345} {\bibfield  {journal} {\bibinfo
  {journal} {Phys. Rev. B}\ }\textbf {\bibinfo {volume} {48}},\ \bibinfo
  {pages} {10345} (\bibinfo {year} {1993})}\BibitemShut {NoStop}%
\bibitem [{\citenamefont {{\"O}stlund}\ and\ \citenamefont
  {Rommer}(1995)}]{OstlundR1995}%
  \BibitemOpen
  \bibfield  {author} {\bibinfo {author} {\bibfnamefont {S.}~\bibnamefont
  {{\"O}stlund}}\ and\ \bibinfo {author} {\bibfnamefont {S.}~\bibnamefont
  {Rommer}},\ }\bibfield  {title} {\bibinfo {title} {Thermodynamic limit of
  density matrix renormalization},\ }\href
  {https://doi.org/10.1103/PhysRevLett.75.3537} {\bibfield  {journal} {\bibinfo
   {journal} {Phys. Rev. Lett.}\ }\textbf {\bibinfo {volume} {75}},\ \bibinfo
  {pages} {3537} (\bibinfo {year} {1995})}\BibitemShut {NoStop}%
\bibitem [{\citenamefont {Rommer}\ and\ \citenamefont
  {{\"O}stlund}(1997)}]{RommerO1997}%
  \BibitemOpen
  \bibfield  {author} {\bibinfo {author} {\bibfnamefont {S.}~\bibnamefont
  {Rommer}}\ and\ \bibinfo {author} {\bibfnamefont {S.}~\bibnamefont
  {{\"O}stlund}},\ }\bibfield  {title} {\bibinfo {title} {Class of ansatz wave
  functions for one-dimensional spin systems and their relation to the density
  matrix renormalization group},\ }\href
  {https://doi.org/10.1103/PhysRevB.55.2164} {\bibfield  {journal} {\bibinfo
  {journal} {Phys. Rev. B}\ }\textbf {\bibinfo {volume} {55}},\ \bibinfo
  {pages} {2164} (\bibinfo {year} {1997})}\BibitemShut {NoStop}%
\bibitem [{\citenamefont {White}\ and\ \citenamefont
  {Feiguin}(2004)}]{WhiteF2004}%
  \BibitemOpen
  \bibfield  {author} {\bibinfo {author} {\bibfnamefont {S.~R.}\ \bibnamefont
  {White}}\ and\ \bibinfo {author} {\bibfnamefont {A.~E.}\ \bibnamefont
  {Feiguin}},\ }\bibfield  {title} {\bibinfo {title} {Real-time evolution using
  the density matrix renormalization group},\ }\href
  {https://doi.org/10.1103/PhysRevLett.93.076401} {\bibfield  {journal}
  {\bibinfo  {journal} {Phys. Rev. Lett.}\ }\textbf {\bibinfo {volume} {93}},\
  \bibinfo {pages} {076401} (\bibinfo {year} {2004})}\BibitemShut {NoStop}%
\bibitem [{\citenamefont {Daley}\ \emph {et~al.}(2004)\citenamefont {Daley},
  \citenamefont {Kollath}, \citenamefont {Schollw\"ock},\ and\ \citenamefont
  {Vidal}}]{DaleyKSV2004}%
  \BibitemOpen
  \bibfield  {author} {\bibinfo {author} {\bibfnamefont {A.~J.}\ \bibnamefont
  {Daley}}, \bibinfo {author} {\bibfnamefont {C.}~\bibnamefont {Kollath}},
  \bibinfo {author} {\bibfnamefont {U.}~\bibnamefont {Schollw\"ock}},\ and\
  \bibinfo {author} {\bibfnamefont {G.}~\bibnamefont {Vidal}},\ }\bibfield
  {title} {\bibinfo {title} {Time-dependent density-matrix
  renormalization-group using adaptive effective hilbert spaces},\ }\href
  {https://doi.org/10.1088/1742-5468/2004/04/P04005} {\bibfield  {journal}
  {\bibinfo  {journal} {Journal of Statistical Mechanics: Theory and
  Experiment}\ }\textbf {\bibinfo {volume} {2004}},\ \bibinfo {pages} {P04005}
  (\bibinfo {year} {2004})}\BibitemShut {NoStop}%
\bibitem [{\citenamefont {Vidal}(2003)}]{Vidal2003}%
  \BibitemOpen
  \bibfield  {author} {\bibinfo {author} {\bibfnamefont {G.}~\bibnamefont
  {Vidal}},\ }\bibfield  {title} {\bibinfo {title} {Efficient classical
  simulation of slightly entangled quantum computations},\ }\href
  {https://doi.org/10.1103/PhysRevLett.91.147902} {\bibfield  {journal}
  {\bibinfo  {journal} {Phys. Rev. Lett.}\ }\textbf {\bibinfo {volume} {91}},\
  \bibinfo {pages} {147902} (\bibinfo {year} {2003})}\BibitemShut {NoStop}%
\bibitem [{\citenamefont {Vidal}(2004)}]{Vidal2004}%
  \BibitemOpen
  \bibfield  {author} {\bibinfo {author} {\bibfnamefont {G.}~\bibnamefont
  {Vidal}},\ }\bibfield  {title} {\bibinfo {title} {Efficient simulation of
  one-dimensional quantum many-body systems},\ }\href
  {https://doi.org/10.1103/PhysRevLett.93.040502} {\bibfield  {journal}
  {\bibinfo  {journal} {Phys. Rev. Lett.}\ }\textbf {\bibinfo {volume} {93}},\
  \bibinfo {pages} {040502} (\bibinfo {year} {2004})}\BibitemShut {NoStop}%
\bibitem [{\citenamefont {White}(2009)}]{White2009}%
  \BibitemOpen
  \bibfield  {author} {\bibinfo {author} {\bibfnamefont {S.~R.}\ \bibnamefont
  {White}},\ }\bibfield  {title} {\bibinfo {title} {Minimally entangled typical
  quantum states at finite temperature},\ }\href
  {https://doi.org/10.1103/PhysRevLett.102.190601} {\bibfield  {journal}
  {\bibinfo  {journal} {Phys. Rev. Lett.}\ }\textbf {\bibinfo {volume} {102}},\
  \bibinfo {pages} {190601} (\bibinfo {year} {2009})}\BibitemShut {NoStop}%
\bibitem [{\citenamefont {Stoudenmire}\ and\ \citenamefont
  {White}(2010)}]{StoudenmireW2010}%
  \BibitemOpen
  \bibfield  {author} {\bibinfo {author} {\bibfnamefont {E.~M.}\ \bibnamefont
  {Stoudenmire}}\ and\ \bibinfo {author} {\bibfnamefont {S.~R.}\ \bibnamefont
  {White}},\ }\bibfield  {title} {\bibinfo {title} {Minimally entangled typical
  thermal state algorithms},\ }\href
  {https://doi.org/10.1088/1367-2630/12/5/055026} {\bibfield  {journal}
  {\bibinfo  {journal} {New Journal of Physics}\ }\textbf {\bibinfo {volume}
  {12}},\ \bibinfo {pages} {055026} (\bibinfo {year} {2010})}\BibitemShut
  {NoStop}%
\bibitem [{\citenamefont {Iwaki}\ \emph {et~al.}(2021)\citenamefont {Iwaki},
  \citenamefont {Shimizu},\ and\ \citenamefont {Hotta}}]{IwakiSH2021}%
  \BibitemOpen
  \bibfield  {author} {\bibinfo {author} {\bibfnamefont {A.}~\bibnamefont
  {Iwaki}}, \bibinfo {author} {\bibfnamefont {A.}~\bibnamefont {Shimizu}},\
  and\ \bibinfo {author} {\bibfnamefont {C.}~\bibnamefont {Hotta}},\ }\bibfield
   {title} {\bibinfo {title} {Thermal pure quantum matrix product states
  recovering a volume law entanglement},\ }\href
  {https://doi.org/10.1103/PhysRevResearch.3.L022015} {\bibfield  {journal}
  {\bibinfo  {journal} {Phys. Rev. Res.}\ }\textbf {\bibinfo {volume} {3}},\
  \bibinfo {pages} {L022015} (\bibinfo {year} {2021})}\BibitemShut {NoStop}%
\bibitem [{\citenamefont {Oseledets}(2011)}]{TT_Review}%
  \BibitemOpen
  \bibfield  {author} {\bibinfo {author} {\bibfnamefont {I.~V.}\ \bibnamefont
  {Oseledets}},\ }\bibfield  {title} {\bibinfo {title} {Tensor-train
  decomposition},\ }\href {https://doi.org/10.1137/090752286} {\bibfield
  {journal} {\bibinfo  {journal} {SIAM Journal on Scientific Computing}\
  }\textbf {\bibinfo {volume} {33}},\ \bibinfo {pages} {2295} (\bibinfo {year}
  {2011})},\ \Eprint {https://arxiv.org/abs/https://doi.org/10.1137/090752286}
  {https://doi.org/10.1137/090752286} \BibitemShut {NoStop}%
\bibitem [{\citenamefont {Stoudenmire}\ and\ \citenamefont
  {Schwab}(2016)}]{MPS_ML}%
  \BibitemOpen
  \bibfield  {author} {\bibinfo {author} {\bibfnamefont {E.}~\bibnamefont
  {Stoudenmire}}\ and\ \bibinfo {author} {\bibfnamefont {D.~J.}\ \bibnamefont
  {Schwab}},\ }\bibfield  {title} {\bibinfo {title} {Supervised learning with
  tensor networks},\ }in\ \href
  {https://proceedings.neurips.cc/paper_files/paper/2016/file/5314b9674c86e3f9d1ba25ef9bb32895-Paper.pdf}
  {\emph {\bibinfo {booktitle} {Advances in Neural Information Processing
  Systems}}},\ Vol.~\bibinfo {volume} {29},\ \bibinfo {editor} {edited by\
  \bibinfo {editor} {\bibfnamefont {D.}~\bibnamefont {Lee}}, \bibinfo {editor}
  {\bibfnamefont {M.}~\bibnamefont {Sugiyama}}, \bibinfo {editor}
  {\bibfnamefont {U.}~\bibnamefont {Luxburg}}, \bibinfo {editor} {\bibfnamefont
  {I.}~\bibnamefont {Guyon}},\ and\ \bibinfo {editor} {\bibfnamefont
  {R.}~\bibnamefont {Garnett}}}\ (\bibinfo  {publisher} {Curran Associates,
  Inc.},\ \bibinfo {year} {2016})\BibitemShut {NoStop}%
\bibitem [{\citenamefont {Novikov}\ \emph {et~al.}(2015)\citenamefont
  {Novikov}, \citenamefont {Podoprikhin}, \citenamefont {Osokin},\ and\
  \citenamefont {Vetrov}}]{TT_DL}%
  \BibitemOpen
  \bibfield  {author} {\bibinfo {author} {\bibfnamefont {A.}~\bibnamefont
  {Novikov}}, \bibinfo {author} {\bibfnamefont {D.}~\bibnamefont
  {Podoprikhin}}, \bibinfo {author} {\bibfnamefont {A.}~\bibnamefont
  {Osokin}},\ and\ \bibinfo {author} {\bibfnamefont {D.~P.}\ \bibnamefont
  {Vetrov}},\ }\bibfield  {title} {\bibinfo {title} {Tensorizing neural
  networks},\ }in\ \href
  {https://proceedings.neurips.cc/paper_files/paper/2015/file/6855456e2fe46a9d49d3d3af4f57443d-Paper.pdf}
  {\emph {\bibinfo {booktitle} {Advances in Neural Information Processing
  Systems}}},\ Vol.~\bibinfo {volume} {28},\ \bibinfo {editor} {edited by\
  \bibinfo {editor} {\bibfnamefont {C.}~\bibnamefont {Cortes}}, \bibinfo
  {editor} {\bibfnamefont {N.}~\bibnamefont {Lawrence}}, \bibinfo {editor}
  {\bibfnamefont {D.}~\bibnamefont {Lee}}, \bibinfo {editor} {\bibfnamefont
  {M.}~\bibnamefont {Sugiyama}},\ and\ \bibinfo {editor} {\bibfnamefont
  {R.}~\bibnamefont {Garnett}}}\ (\bibinfo  {publisher} {Curran Associates,
  Inc.},\ \bibinfo {year} {2015})\BibitemShut {NoStop}%
\bibitem [{\citenamefont {Shinaoka}\ \emph {et~al.}(2023)\citenamefont
  {Shinaoka}, \citenamefont {Wallerberger}, \citenamefont {Murakami},
  \citenamefont {Nogaki}, \citenamefont {Sakurai}, \citenamefont {Werner},\
  and\ \citenamefont {Kauch}}]{Shinaokaetal2023}%
  \BibitemOpen
  \bibfield  {author} {\bibinfo {author} {\bibfnamefont {H.}~\bibnamefont
  {Shinaoka}}, \bibinfo {author} {\bibfnamefont {M.}~\bibnamefont
  {Wallerberger}}, \bibinfo {author} {\bibfnamefont {Y.}~\bibnamefont
  {Murakami}}, \bibinfo {author} {\bibfnamefont {K.}~\bibnamefont {Nogaki}},
  \bibinfo {author} {\bibfnamefont {R.}~\bibnamefont {Sakurai}}, \bibinfo
  {author} {\bibfnamefont {P.}~\bibnamefont {Werner}},\ and\ \bibinfo {author}
  {\bibfnamefont {A.}~\bibnamefont {Kauch}},\ }\bibfield  {title} {\bibinfo
  {title} {Multiscale space-time ansatz for correlation functions of quantum
  systems based on quantics tensor trains},\ }\href
  {https://doi.org/10.1103/PhysRevX.13.021015} {\bibfield  {journal} {\bibinfo
  {journal} {Phys. Rev. X}\ }\textbf {\bibinfo {volume} {13}},\ \bibinfo
  {pages} {021015} (\bibinfo {year} {2023})}\BibitemShut {NoStop}%
\bibitem [{\citenamefont {Fishman}\ \emph
  {et~al.}(2022{\natexlab{a}})\citenamefont {Fishman}, \citenamefont {White},\
  and\ \citenamefont {Stoudenmire}}]{iTensor}%
  \BibitemOpen
  \bibfield  {author} {\bibinfo {author} {\bibfnamefont {M.}~\bibnamefont
  {Fishman}}, \bibinfo {author} {\bibfnamefont {S.~R.}\ \bibnamefont {White}},\
  and\ \bibinfo {author} {\bibfnamefont {E.~M.}\ \bibnamefont {Stoudenmire}},\
  }\bibfield  {title} {\bibinfo {title} {{The ITensor Software Library for
  Tensor Network Calculations}},\ }\href
  {https://doi.org/10.21468/SciPostPhysCodeb.4} {\bibfield  {journal} {\bibinfo
   {journal} {SciPost Phys. Codebases}\ ,\ \bibinfo {pages} {4}} (\bibinfo
  {year} {2022}{\natexlab{a}})}\BibitemShut {NoStop}%
\bibitem [{\citenamefont {Fishman}\ \emph
  {et~al.}(2022{\natexlab{b}})\citenamefont {Fishman}, \citenamefont {White},\
  and\ \citenamefont {Stoudenmire}}]{iTensor03}%
  \BibitemOpen
  \bibfield  {author} {\bibinfo {author} {\bibfnamefont {M.}~\bibnamefont
  {Fishman}}, \bibinfo {author} {\bibfnamefont {S.~R.}\ \bibnamefont {White}},\
  and\ \bibinfo {author} {\bibfnamefont {E.~M.}\ \bibnamefont {Stoudenmire}},\
  }\bibfield  {title} {\bibinfo {title} {{Codebase release 0.3 for ITensor}},\
  }\href {https://doi.org/10.21468/SciPostPhysCodeb.4-r0.3} {\bibfield
  {journal} {\bibinfo  {journal} {SciPost Phys. Codebases}\ ,\ \bibinfo {pages}
  {4}} (\bibinfo {year} {2022}{\natexlab{b}})}\BibitemShut {NoStop}%
\bibitem [{\citenamefont {Zhai}\ \emph {et~al.}(2023)\citenamefont {Zhai},
  \citenamefont {Larsson}, \citenamefont {Lee}, \citenamefont {Cui},
  \citenamefont {Zhu}, \citenamefont {Sun}, \citenamefont {Peng}, \citenamefont
  {Peng}, \citenamefont {Liao}, \citenamefont {T\"olle}, \citenamefont {Yang},
  \citenamefont {Li},\ and\ \citenamefont {Chan}}]{10.1063/5.0180424}%
  \BibitemOpen
  \bibfield  {author} {\bibinfo {author} {\bibfnamefont {H.}~\bibnamefont
  {Zhai}}, \bibinfo {author} {\bibfnamefont {H.~R.}\ \bibnamefont {Larsson}},
  \bibinfo {author} {\bibfnamefont {S.}~\bibnamefont {Lee}}, \bibinfo {author}
  {\bibfnamefont {Z.-H.}\ \bibnamefont {Cui}}, \bibinfo {author} {\bibfnamefont
  {T.}~\bibnamefont {Zhu}}, \bibinfo {author} {\bibfnamefont {C.}~\bibnamefont
  {Sun}}, \bibinfo {author} {\bibfnamefont {L.}~\bibnamefont {Peng}}, \bibinfo
  {author} {\bibfnamefont {R.}~\bibnamefont {Peng}}, \bibinfo {author}
  {\bibfnamefont {K.}~\bibnamefont {Liao}}, \bibinfo {author} {\bibfnamefont
  {J.}~\bibnamefont {T\"olle}}, \bibinfo {author} {\bibfnamefont
  {J.}~\bibnamefont {Yang}}, \bibinfo {author} {\bibfnamefont {S.}~\bibnamefont
  {Li}},\ and\ \bibinfo {author} {\bibfnamefont {G.~K.-L.}\ \bibnamefont
  {Chan}},\ }\bibfield  {title} {\bibinfo {title} {Block2: A comprehensive open
  source framework to develop and apply state-of-the-art dmrg algorithms in
  electronic structure and beyond},\ }\href {https://doi.org/10.1063/5.0180424}
  {\bibfield  {journal} {\bibinfo  {journal} {The Journal of Chemical Physics}\
  }\textbf {\bibinfo {volume} {159}},\ \bibinfo {pages} {234801} (\bibinfo
  {year} {2023})}\BibitemShut {NoStop}%
\bibitem [{\citenamefont {Hauschild}\ \emph {et~al.}(2024)\citenamefont
  {Hauschild}, \citenamefont {Unfried}, \citenamefont {Anand}, \citenamefont
  {Andrews}, \citenamefont {Bintz}, \citenamefont {Borla}, \citenamefont
  {Divic}, \citenamefont {Drescher}, \citenamefont {Geiger}, \citenamefont
  {Hefel}, \citenamefont {H\'emery}, \citenamefont {Kadow}, \citenamefont
  {Kemp}, \citenamefont {Kirchner}, \citenamefont {Liu}, \citenamefont
  {M\"oller}, \citenamefont {Parker}, \citenamefont {Rader}, \citenamefont
  {Romen}, \citenamefont {Scalet}, \citenamefont {Schoonderwoerd},
  \citenamefont {Schulz}, \citenamefont {Soejima}, \citenamefont {Thoma},
  \citenamefont {Wu}, \citenamefont {Zechmann}, \citenamefont {Zweng},
  \citenamefont {Mong}, \citenamefont {Zaletel},\ and\ \citenamefont
  {Pollmann}}]{tenpy2024}%
  \BibitemOpen
  \bibfield  {author} {\bibinfo {author} {\bibfnamefont {J.}~\bibnamefont
  {Hauschild}}, \bibinfo {author} {\bibfnamefont {J.}~\bibnamefont {Unfried}},
  \bibinfo {author} {\bibfnamefont {S.}~\bibnamefont {Anand}}, \bibinfo
  {author} {\bibfnamefont {B.}~\bibnamefont {Andrews}}, \bibinfo {author}
  {\bibfnamefont {M.}~\bibnamefont {Bintz}}, \bibinfo {author} {\bibfnamefont
  {U.}~\bibnamefont {Borla}}, \bibinfo {author} {\bibfnamefont
  {S.}~\bibnamefont {Divic}}, \bibinfo {author} {\bibfnamefont
  {M.}~\bibnamefont {Drescher}}, \bibinfo {author} {\bibfnamefont
  {J.}~\bibnamefont {Geiger}}, \bibinfo {author} {\bibfnamefont
  {M.}~\bibnamefont {Hefel}}, \bibinfo {author} {\bibfnamefont
  {K.}~\bibnamefont {H\'emery}}, \bibinfo {author} {\bibfnamefont
  {W.}~\bibnamefont {Kadow}}, \bibinfo {author} {\bibfnamefont
  {J.}~\bibnamefont {Kemp}}, \bibinfo {author} {\bibfnamefont {N.}~\bibnamefont
  {Kirchner}}, \bibinfo {author} {\bibfnamefont {V.~S.}\ \bibnamefont {Liu}},
  \bibinfo {author} {\bibfnamefont {G.}~\bibnamefont {M\"oller}}, \bibinfo
  {author} {\bibfnamefont {D.}~\bibnamefont {Parker}}, \bibinfo {author}
  {\bibfnamefont {M.}~\bibnamefont {Rader}}, \bibinfo {author} {\bibfnamefont
  {A.}~\bibnamefont {Romen}}, \bibinfo {author} {\bibfnamefont
  {S.}~\bibnamefont {Scalet}}, \bibinfo {author} {\bibfnamefont
  {L.}~\bibnamefont {Schoonderwoerd}}, \bibinfo {author} {\bibfnamefont
  {M.}~\bibnamefont {Schulz}}, \bibinfo {author} {\bibfnamefont
  {T.}~\bibnamefont {Soejima}}, \bibinfo {author} {\bibfnamefont
  {P.}~\bibnamefont {Thoma}}, \bibinfo {author} {\bibfnamefont
  {Y.}~\bibnamefont {Wu}}, \bibinfo {author} {\bibfnamefont {P.}~\bibnamefont
  {Zechmann}}, \bibinfo {author} {\bibfnamefont {L.}~\bibnamefont {Zweng}},
  \bibinfo {author} {\bibfnamefont {R.~S.~K.}\ \bibnamefont {Mong}}, \bibinfo
  {author} {\bibfnamefont {M.~P.}\ \bibnamefont {Zaletel}},\ and\ \bibinfo
  {author} {\bibfnamefont {F.}~\bibnamefont {Pollmann}},\ }\bibfield  {title}
  {\bibinfo {title} {{Tensor network Python (TeNPy) version 1}},\ }\href
  {https://doi.org/10.21468/SciPostPhysCodeb.41} {\bibfield  {journal}
  {\bibinfo  {journal} {SciPost Phys. Codebases}\ ,\ \bibinfo {pages} {41}}
  (\bibinfo {year} {2024})}\BibitemShut {NoStop}%
\bibitem [{\citenamefont {Van~Damme}\ \emph {et~al.}(2025)\citenamefont
  {Van~Damme}, \citenamefont {Devos},\ and\ \citenamefont
  {Haegeman}}]{van_damme_2025_17313329}%
  \BibitemOpen
  \bibfield  {author} {\bibinfo {author} {\bibfnamefont {M.}~\bibnamefont
  {Van~Damme}}, \bibinfo {author} {\bibfnamefont {L.}~\bibnamefont {Devos}},\
  and\ \bibinfo {author} {\bibfnamefont {J.}~\bibnamefont {Haegeman}},\ }\href
  {https://doi.org/10.5281/zenodo.17313329} {\bibinfo {title} {Mpskit}}
  (\bibinfo {year} {2025})\BibitemShut {NoStop}%
\bibitem [{\citenamefont {Chan}\ and\ \citenamefont
  {Head-Gordon}(2002)}]{ChanHG2002}%
  \BibitemOpen
  \bibfield  {author} {\bibinfo {author} {\bibfnamefont {G.~K.-L.}\
  \bibnamefont {Chan}}\ and\ \bibinfo {author} {\bibfnamefont {M.}~\bibnamefont
  {Head-Gordon}},\ }\bibfield  {title} {\bibinfo {title} {{Highly correlated
  calculations with a polynomial cost algorithm: A study of the density matrix
  renormalization group}},\ }\href {https://doi.org/10.1063/1.1449459}
  {\bibfield  {journal} {\bibinfo  {journal} {The Journal of Chemical Physics}\
  }\textbf {\bibinfo {volume} {116}},\ \bibinfo {pages} {4462} (\bibinfo {year}
  {2002})},\ \Eprint
  {https://arxiv.org/abs/https://pubs.aip.org/aip/jcp/article-pdf/116/11/4462/19222618/4462\_1\_online.pdf}
  {https://pubs.aip.org/aip/jcp/article-pdf/116/11/4462/19222618/4462\_1\_online.pdf}
  \BibitemShut {NoStop}%
\bibitem [{\citenamefont {Legeza}\ and\ \citenamefont
  {S\'olyom}(2003)}]{LegezaS2003}%
  \BibitemOpen
  \bibfield  {author} {\bibinfo {author} {\bibfnamefont {O.}~\bibnamefont
  {Legeza}}\ and\ \bibinfo {author} {\bibfnamefont {J.}~\bibnamefont
  {S\'olyom}},\ }\bibfield  {title} {\bibinfo {title} {Optimizing the
  density-matrix renormalization group method using quantum information
  entropy},\ }\href {https://doi.org/10.1103/PhysRevB.68.195116} {\bibfield
  {journal} {\bibinfo  {journal} {Phys. Rev. B}\ }\textbf {\bibinfo {volume}
  {68}},\ \bibinfo {pages} {195116} (\bibinfo {year} {2003})}\BibitemShut
  {NoStop}%
\bibitem [{\citenamefont {Moritz}\ \emph {et~al.}(2004)\citenamefont {Moritz},
  \citenamefont {Hess},\ and\ \citenamefont {Reiher}}]{MoritzHR2004}%
  \BibitemOpen
  \bibfield  {author} {\bibinfo {author} {\bibfnamefont {G.}~\bibnamefont
  {Moritz}}, \bibinfo {author} {\bibfnamefont {B.~A.}\ \bibnamefont {Hess}},\
  and\ \bibinfo {author} {\bibfnamefont {M.}~\bibnamefont {Reiher}},\
  }\bibfield  {title} {\bibinfo {title} {{Convergence behavior of the
  density-matrix renormalization group algorithm for optimized orbital
  orderings}},\ }\href {https://doi.org/10.1063/1.1824891} {\bibfield
  {journal} {\bibinfo  {journal} {The Journal of Chemical Physics}\ }\textbf
  {\bibinfo {volume} {122}},\ \bibinfo {pages} {024107} (\bibinfo {year}
  {2004})},\ \Eprint
  {https://arxiv.org/abs/https://pubs.aip.org/aip/jcp/article-pdf/doi/10.1063/1.1824891/10872404/024107\_1\_online.pdf}
  {https://pubs.aip.org/aip/jcp/article-pdf/doi/10.1063/1.1824891/10872404/024107\_1\_online.pdf}
  \BibitemShut {NoStop}%
\bibitem [{\citenamefont {Legeza}\ \emph {et~al.}(2015)\citenamefont {Legeza},
  \citenamefont {Veis}, \citenamefont {Poves},\ and\ \citenamefont
  {Dukelsky}}]{LegezaVPD2015}%
  \BibitemOpen
  \bibfield  {author} {\bibinfo {author} {\bibfnamefont {O.}~\bibnamefont
  {Legeza}}, \bibinfo {author} {\bibfnamefont {L.}~\bibnamefont {Veis}},
  \bibinfo {author} {\bibfnamefont {A.}~\bibnamefont {Poves}},\ and\ \bibinfo
  {author} {\bibfnamefont {J.}~\bibnamefont {Dukelsky}},\ }\bibfield  {title}
  {\bibinfo {title} {Advanced density matrix renormalization group method for
  nuclear structure calculations},\ }\href
  {https://doi.org/10.1103/PhysRevC.92.051303} {\bibfield  {journal} {\bibinfo
  {journal} {Phys. Rev. C}\ }\textbf {\bibinfo {volume} {92}},\ \bibinfo
  {pages} {051303} (\bibinfo {year} {2015})}\BibitemShut {NoStop}%
\bibitem [{\citenamefont {Li}\ \emph {et~al.}(2022)\citenamefont {Li},
  \citenamefont {Ren}, \citenamefont {Yang},\ and\ \citenamefont
  {Shuai}}]{LiRYS2022}%
  \BibitemOpen
  \bibfield  {author} {\bibinfo {author} {\bibfnamefont {W.}~\bibnamefont
  {Li}}, \bibinfo {author} {\bibfnamefont {J.}~\bibnamefont {Ren}}, \bibinfo
  {author} {\bibfnamefont {H.}~\bibnamefont {Yang}},\ and\ \bibinfo {author}
  {\bibfnamefont {Z.}~\bibnamefont {Shuai}},\ }\bibfield  {title} {\bibinfo
  {title} {On the fly swapping algorithm for ordering of degrees of freedom in
  density matrix renormalization group},\ }\href
  {https://doi.org/10.1088/1361-648X/ac640e} {\bibfield  {journal} {\bibinfo
  {journal} {Journal of Physics: Condensed Matter}\ }\textbf {\bibinfo {volume}
  {34}},\ \bibinfo {pages} {254003} (\bibinfo {year} {2022})}\BibitemShut
  {NoStop}%
\bibitem [{\citenamefont {Larsson}(2019)}]{Larsson2019}%
  \BibitemOpen
  \bibfield  {author} {\bibinfo {author} {\bibfnamefont {H.~R.}\ \bibnamefont
  {Larsson}},\ }\bibfield  {title} {\bibinfo {title} {Computing vibrational
  eigenstates with tree tensor network states (ttns)},\ }\href
  {https://doi.org/10.1063/1.5130390} {\bibfield  {journal} {\bibinfo
  {journal} {The Journal of Chemical Physics}\ }\textbf {\bibinfo {volume}
  {151}},\ \bibinfo {pages} {204102} (\bibinfo {year} {2019})},\ \Eprint
  {https://arxiv.org/abs/https://doi.org/10.1063/1.5130390}
  {https://doi.org/10.1063/1.5130390} \BibitemShut {NoStop}%
\bibitem [{\citenamefont {Hikihara}\ \emph {et~al.}(2023)\citenamefont
  {Hikihara}, \citenamefont {Ueda}, \citenamefont {Okunishi}, \citenamefont
  {Harada},\ and\ \citenamefont {Nishino}}]{HikiharaUOHN2023}%
  \BibitemOpen
  \bibfield  {author} {\bibinfo {author} {\bibfnamefont {T.}~\bibnamefont
  {Hikihara}}, \bibinfo {author} {\bibfnamefont {H.}~\bibnamefont {Ueda}},
  \bibinfo {author} {\bibfnamefont {K.}~\bibnamefont {Okunishi}}, \bibinfo
  {author} {\bibfnamefont {K.}~\bibnamefont {Harada}},\ and\ \bibinfo {author}
  {\bibfnamefont {T.}~\bibnamefont {Nishino}},\ }\bibfield  {title} {\bibinfo
  {title} {Automatic structural optimization of tree tensor networks},\ }\href
  {https://doi.org/10.1103/PhysRevResearch.5.013031} {\bibfield  {journal}
  {\bibinfo  {journal} {Phys. Rev. Res.}\ }\textbf {\bibinfo {volume} {5}},\
  \bibinfo {pages} {013031} (\bibinfo {year} {2023})}\BibitemShut {NoStop}%
\bibitem [{\citenamefont {Hikihara}\ \emph {et~al.}(2025)\citenamefont
  {Hikihara}, \citenamefont {Ueda}, \citenamefont {Okunishi}, \citenamefont
  {Harada},\ and\ \citenamefont {Nishino}}]{HikiharaUOHN2025}%
  \BibitemOpen
  \bibfield  {author} {\bibinfo {author} {\bibfnamefont {T.}~\bibnamefont
  {Hikihara}}, \bibinfo {author} {\bibfnamefont {H.}~\bibnamefont {Ueda}},
  \bibinfo {author} {\bibfnamefont {K.}~\bibnamefont {Okunishi}}, \bibinfo
  {author} {\bibfnamefont {K.}~\bibnamefont {Harada}},\ and\ \bibinfo {author}
  {\bibfnamefont {T.}~\bibnamefont {Nishino}},\ }\bibfield  {title} {\bibinfo
  {title} {Improving the accuracy of the tree-tensor network approach by
  optimization of network structure},\ }\href
  {https://doi.org/10.1103/ljj8-tkpc} {\bibfield  {journal} {\bibinfo
  {journal} {Phys. Rev. B}\ }\textbf {\bibinfo {volume} {112}},\ \bibinfo
  {pages} {134427} (\bibinfo {year} {2025})}\BibitemShut {NoStop}%
\bibitem [{\citenamefont {Watanabe}\ \emph {et~al.}(2026)\citenamefont
  {Watanabe}, \citenamefont {Manabe}, \citenamefont {Hikihara},\ and\
  \citenamefont {Ueda}}]{WatanabeMHU2026}%
  \BibitemOpen
  \bibfield  {author} {\bibinfo {author} {\bibfnamefont {R.}~\bibnamefont
  {Watanabe}}, \bibinfo {author} {\bibfnamefont {H.}~\bibnamefont {Manabe}},
  \bibinfo {author} {\bibfnamefont {T.}~\bibnamefont {Hikihara}},\ and\
  \bibinfo {author} {\bibfnamefont {H.}~\bibnamefont {Ueda}},\ }\bibfield
  {title} {\bibinfo {title} {Ttnopt: Tree tensor network package for high-rank
  tensor compression},\ }\href
  {https://doi.org/https://doi.org/10.1016/j.cpc.2025.109918} {\bibfield
  {journal} {\bibinfo  {journal} {Computer Physics Communications}\ }\textbf
  {\bibinfo {volume} {319}},\ \bibinfo {pages} {109918} (\bibinfo {year}
  {2026})}\BibitemShut {NoStop}%
\bibitem [{not()}]{note_EEreduction}%
  \BibitemOpen
  \bibinfo {note} {To be precise, entanglement entropy may slightly decrease by
  the two-site swap also in this case, since it is not precisely proportional
  to the number of blue lines. However, even in such a case, the approach
  toward the optimal site order may be hindered.}\BibitemShut {Stop}%
\bibitem [{\citenamefont {Shi}\ \emph {et~al.}(2006)\citenamefont {Shi},
  \citenamefont {Duan},\ and\ \citenamefont {Vidal}}]{ShiDV2006}%
  \BibitemOpen
  \bibfield  {author} {\bibinfo {author} {\bibfnamefont {Y.-Y.}\ \bibnamefont
  {Shi}}, \bibinfo {author} {\bibfnamefont {L.-M.}\ \bibnamefont {Duan}},\ and\
  \bibinfo {author} {\bibfnamefont {G.}~\bibnamefont {Vidal}},\ }\bibfield
  {title} {\bibinfo {title} {Classical simulation of quantum many-body systems
  with a tree tensor network},\ }\href
  {https://doi.org/10.1103/PhysRevA.74.022320} {\bibfield  {journal} {\bibinfo
  {journal} {Phys. Rev. A}\ }\textbf {\bibinfo {volume} {74}},\ \bibinfo
  {pages} {022320} (\bibinfo {year} {2006})}\BibitemShut {NoStop}%
\bibitem [{con()}]{convergence}%
  \BibitemOpen
  \bibinfo {note} {In practice, if the calculation does not converge even after
  100 iterations of the processes of the first and second parts, we have
  terminated the calculation and adopted the site order obtained at that point
  as the optimized one. The number of such non-converged calculations was 12
  out of the site order optimization calculations performed for 2500 parameter
  sets.}\BibitemShut {Stop}%
\bibitem [{\citenamefont {Watanabe}\ and\ \citenamefont
  {Shinaoka}(2025)}]{TreeTCI}%
  \BibitemOpen
  \bibfield  {author} {\bibinfo {author} {\bibfnamefont {R.}~\bibnamefont
  {Watanabe}}\ and\ \bibinfo {author} {\bibfnamefont {H.}~\bibnamefont
  {Shinaoka}},\ }\href {https://github.com/tensor4all/TreeTCI.jl} {\bibinfo
  {title} {{Tree}{TCI}}},\ \bibinfo {howpublished}
  {\url{https://github.com/tensor4all/TreeTCI.jl}} (\bibinfo {year}
  {2025})\BibitemShut {NoStop}%
\bibitem [{\citenamefont {Fern\'andez}\ \emph {et~al.}(2025)\citenamefont
  {Fern\'andez}, \citenamefont {Ritter}, \citenamefont {Jeannin}, \citenamefont
  {Li}, \citenamefont {Kloss}, \citenamefont {Louvet}, \citenamefont
  {Terasaki}, \citenamefont {Parcollet}, \citenamefont {von Delft},
  \citenamefont {Shinaoka},\ and\ \citenamefont {Waintal}}]{Nunezetal2025}%
  \BibitemOpen
  \bibfield  {author} {\bibinfo {author} {\bibfnamefont {Y.~N.}\ \bibnamefont
  {Fern\'andez}}, \bibinfo {author} {\bibfnamefont {M.~K.}\ \bibnamefont
  {Ritter}}, \bibinfo {author} {\bibfnamefont {M.}~\bibnamefont {Jeannin}},
  \bibinfo {author} {\bibfnamefont {J.-W.}\ \bibnamefont {Li}}, \bibinfo
  {author} {\bibfnamefont {T.}~\bibnamefont {Kloss}}, \bibinfo {author}
  {\bibfnamefont {T.}~\bibnamefont {Louvet}}, \bibinfo {author} {\bibfnamefont
  {S.}~\bibnamefont {Terasaki}}, \bibinfo {author} {\bibfnamefont
  {O.}~\bibnamefont {Parcollet}}, \bibinfo {author} {\bibfnamefont
  {J.}~\bibnamefont {von Delft}}, \bibinfo {author} {\bibfnamefont
  {H.}~\bibnamefont {Shinaoka}},\ and\ \bibinfo {author} {\bibfnamefont
  {X.}~\bibnamefont {Waintal}},\ }\bibfield  {title} {\bibinfo {title}
  {{Learning tensor networks with tensor cross interpolation: New algorithms
  and libraries}},\ }\href {https://doi.org/10.21468/SciPostPhys.18.3.104}
  {\bibfield  {journal} {\bibinfo  {journal} {SciPost Phys.}\ }\textbf
  {\bibinfo {volume} {18}},\ \bibinfo {pages} {104} (\bibinfo {year}
  {2025})}\BibitemShut {NoStop}%
\bibitem [{\citenamefont {Qian}\ and\ \citenamefont {Qin}(2023)}]{Qian_2023}%
  \BibitemOpen
  \bibfield  {author} {\bibinfo {author} {\bibfnamefont {X.}~\bibnamefont
  {Qian}}\ and\ \bibinfo {author} {\bibfnamefont {M.}~\bibnamefont {Qin}},\
  }\bibfield  {title} {\bibinfo {title} {Augmenting density matrix
  renormalization group with disentanglers},\ }\href
  {https://doi.org/10.1088/0256-307x/40/5/057102} {\bibfield  {journal}
  {\bibinfo  {journal} {Chinese Physics Letters}\ }\textbf {\bibinfo {volume}
  {40}},\ \bibinfo {pages} {057102} (\bibinfo {year} {2023})}\BibitemShut
  {NoStop}%
\bibitem [{\citenamefont {Qian}\ \emph {et~al.}(2024)\citenamefont {Qian},
  \citenamefont {Huang},\ and\ \citenamefont {Qin}}]{PhysRevLett.133.190402}%
  \BibitemOpen
  \bibfield  {author} {\bibinfo {author} {\bibfnamefont {X.}~\bibnamefont
  {Qian}}, \bibinfo {author} {\bibfnamefont {J.}~\bibnamefont {Huang}},\ and\
  \bibinfo {author} {\bibfnamefont {M.}~\bibnamefont {Qin}},\ }\bibfield
  {title} {\bibinfo {title} {Augmenting density matrix renormalization group
  with clifford circuits},\ }\href
  {https://doi.org/10.1103/PhysRevLett.133.190402} {\bibfield  {journal}
  {\bibinfo  {journal} {Phys. Rev. Lett.}\ }\textbf {\bibinfo {volume} {133}},\
  \bibinfo {pages} {190402} (\bibinfo {year} {2024})}\BibitemShut {NoStop}%
\bibitem [{\citenamefont {Huang}\ \emph {et~al.}(2025)\citenamefont {Huang},
  \citenamefont {Qian}, \citenamefont {Li},\ and\ \citenamefont
  {Qin}}]{huang2025augmentingdensitymatrixrenormalization}%
  \BibitemOpen
  \bibfield  {author} {\bibinfo {author} {\bibfnamefont {J.}~\bibnamefont
  {Huang}}, \bibinfo {author} {\bibfnamefont {X.}~\bibnamefont {Qian}},
  \bibinfo {author} {\bibfnamefont {Z.}~\bibnamefont {Li}},\ and\ \bibinfo
  {author} {\bibfnamefont {M.}~\bibnamefont {Qin}},\ }\href
  {https://arxiv.org/abs/2505.08635} {\bibinfo {title} {Augmenting density
  matrix renormalization group with matchgates and clifford circuits}}
  (\bibinfo {year} {2025}),\ \Eprint {https://arxiv.org/abs/2505.08635}
  {arXiv:2505.08635 [quant-ph]} \BibitemShut {NoStop}%
\bibitem [{\citenamefont {Watanabe}(2025)}]{FlexibleDMRG}%
  \BibitemOpen
  \bibfield  {author} {\bibinfo {author} {\bibfnamefont {R.}~\bibnamefont
  {Watanabe}},\ }\href {https://github.com/Ryo-wtnb11/FlexibleDMRG} {\bibinfo
  {title} {{Flexible}{DMRG}}},\ \bibinfo {howpublished}
  {\url{https://github.com/Ryo-wtnb11/FlexibleDMRG}} (\bibinfo {year}
  {2025})\BibitemShut {NoStop}%
\end{thebibliography}
%

\end{document}